\documentclass[fleqn,11pt]{wlscirep}
\title{Prediction of the Strength and Timing of Sunspot Cycle 25 Reveal Decadal-scale Space Environmental Conditions}

\author[1]{Prantika Bhowmik}
\author[1,2,*]{Dibyendu Nandy}
\affil[1]{Center of Excellence in Space Sciences India, Indian Institute of Science Education and Research Kolkata, Mohanpur 741246, West Bengal, India}
\affil[2]{Department of Physical Sciences, Indian Institute of Science Education and Research Kolkata, Mohanpur 741246, West Bengal, India}

\affil[*]{Corresponding author: dnandi@iiserkol.ac.in}

\begin{abstract}
The Sun's activity cycle governs the radiation, particle and magnetic flux in the heliosphere creating hazardous space weather. Decadal-scale variations define space climate and force the Earth's atmosphere. However, predicting the solar cycle is challenging. Current understanding indicates a short window for prediction best achieved at previous cycle minima. Utilizing magnetic field evolution models for the Sun's surface and interior we perform the first century-scale, data-driven simulations of solar activity and present a scheme for extending the prediction window to a decade. Our ensemble forecast indicates cycle 25 would be similar or slightly stronger than the current cycle and peak around 2024. Sunspot cycle 25 may thus reverse the substantial weakening trend in solar activity which has led to speculation of an imminent Maunder-like grand minimum and cooling global climate. Our simulations demonstrate fluctuation in the tilt angle distribution of sunspots is the dominant mechanism responsible for solar cycle variability.

\end{abstract}
\begin{document}

\flushbottom
\maketitle

\thispagestyle{empty}

\noindent Understanding magnetic field generation in the Sun and stars is an outstanding challenge in astrophysics. Theoretical advances in the context of the solar cycle provide a window to the magnetic Universe on the one hand, and on the other, benefits the quest for predicting space weather and climate. The 11-year cycle of sunspots spawn severe space weather characterized by solar flares, coronal mass ejections, geomagnetic storms, enhanced radiative and energetic particle flux endangering satellites, global communication systems, air-traffic over polar routes and electric power grids\cite{2015AdSpR..55.2745S}. Protection of planetary technologies and space situational awareness is therefore enabled by solar activity predictions. Slow long-term changes in the Sun's radiative energy output -- which is governed by its magnetic activity -- is the primary (external) natural driver of planetary atmospheric dynamics, including climate. Assessment of the strength of future sunspot cycles and its expected radiative output provides critical inputs to climate assessment models\cite{2005SSRv..120..243V,IPCC2014}. 

Sunspots have been observed for over four centuries, constituting the longest running, continuous time series of any natural phenomena in the Universe. However, the fact that they are magnetic in nature and their time-variation a manifestation of an underlying magnetic cycle, has only been known since the beginning of 20$^\textnormal{th}$ century when G. E. Hale and his collaborators discovered that sunspots are strongly magnetized\cite{1908ApJ....28..315H}. They also discovered that sunspots typically appear in pairs with a leading (in the direction of solar rotation) and a following polarity of opposite magnetic signs\cite{1919ApJ....49..153H}. For an individual sunspot cycle, the leading polarities of these Bipolar Magnetic Regions (BMRs) have opposite signs in the two hemispheres. This relative polarity orientation flips from one sunspot cycle to another generating a 22-year magnetic cycle. The amplitude of the sunspot cycle itself -- which determines its space weather consequences -- is highly variable and difficult to predict.

The sunspot cycle is understood to originate via a magnetohydrodynamic (MHD) dynamo mechanism\cite{2010LRSP....7....3C} involving complex interactions between plasma flows and magnetic fields in the solar convection zone (SCZ). Differential rotation in the solar interior stretches the large-scale poloidal component of the Sun's magnetic field in the $\phi$-direction to produce the toroidal component\cite{1955ApJ...122..293P}. Strong toroidal flux tubes rise through the SCZ due to magnetic buoyancy and appear as BMRs on the Sun's surface -- the photosphere. These BMRs are tilted because of the action of the Coriolis force on rising magnetic flux tubes and there is a dispersion observed around the mean tilt which is thought to be due to random buffeting of the rising flux tubes by turbulent convection. It is now believed that the dispersal and decay of these tilted BMRs -- facilitated by surface flux transport processes -- is the predominant mechanism for the regeneration of the Sun's poloidal component\cite{1961ApJ...133..572B,1969ApJ...156....1L,2010A&A...518A...7D,2015Sci...347.1333C}. The latter in turn seeds the generation of the next cycle toroidal component, thereby, sustaining the solar magnetic cycle. An alternate mechanism termed as the mean-field $\alpha$-effect driven by the action of helical turbulent convection on weaker toroidal fields is also thought to contribute to the poloidal feld creation process\cite{2010LRSP....7....3C}. Fluctuations in these poloidal field creation mechanism(s) -- due to the turbulent nature of the solar convection zone -- are likely candidates for governing solar activity variations. 
 
Appropriately constrained computational solar dynamo models, driven by observations, are expected to serve as useful tools for forecasting the sunspot cycle. However, this has remained a challenging task and it has been argued that long-term solar cycle forecasts are not possible\cite{2007ApJ...661.1289B}. Indeed multiple forecasts were made for the current solar cycle 24 with little consensus\cite{2008SoPh..252..209P} and two solar dynamo-based forecasts for cycle 24 differed significantly from each other\cite{2006GeoRL..33.5102D,2007PhRvL..98m1103C,2007MNRAS.381.1527J}. In this backdrop, recent progress focussed on understanding the physics of solar cycle predictability has underscored the importance of plasma flux transport processes in governing the underlying dynamical memory leading to solar cycle predictability, reconciled the difference between diverging dynamo-based forecasts for the current cycle and indicated that the predictive window based on dynamo models alone, is, in fact, short\cite{2008ApJ...673..544Y,2010LRSP....7....6P,2012ApJ...761L..13K}. These and other studies\cite{1999JGR...10422375H,2005GeoRL..32.1104S,2009ApJ...694L..11W,2013ApJ...767L..25M} indicate that prediction of the strength of the next sunspot cycle is indeed plausible, and is best achieved with accurate knowledge (i.e., observational input) of the solar polar (poloidal) field proxy at the preceding cycle minimum, i.e., only about 5 years in advance.

Can we extend this prediction window further? Here we demonstrate that this is viable. We devise a novel methodology, wherein, we first predict the strength of the Sun's polar field at cycle minimum (in advance) and then utilize this as input in a predictive dynamo model to forecast the strength and timing of the next sunspot cycle thereby extending the prediction window to close to a decade.

The transport and dissipation of photospheric magnetic fields that lead to solar polar field reversal and build-up -- peaking at the solar minimum -- is a complex process. The large-scale behavior of the surface magnetic field was first explained by Leighton\cite{1969ApJ...156....1L} who suggested that the magnetic field associated with BMRs diffuses due to a random-walk-like movement of the supergranular convective cells. This diffusion results in flux cancelation along the equator between the leading polarities of BMRs belonging to different hemispheres. As an outcome, there is an imbalance of signed magnetic flux in each solar hemisphere. This excess flux from the following polarities eventually migrate towards the poles and cancels and reverses the old solar cycle polarity. This polar field is, in fact, the radial component of the Sun's poloidal field. Differential rotation and a large-scale flow of plasma from the Sun's equator to the poles known as meridional circulation play crucial roles in this process. This understanding has led to the development of solar Surface Flux Transport (SFT) models which can reasonably simulate the surface dynamics of solar magnetic fields\cite{1989Sci...245..712W,1998ApJ...501..866V,2001ApJ...547..475S,2005LRSP....2....5S,2010ApJ...719..264C,2012LRSP....9....6M,2014SSRv..186..491J}. Earlier simulations with such models indicate a polar field strength at cycle 24 minimum which is weaker or comparable to the previous cycle minimum\cite{2016JGRA..12110744H,2016ApJ...823L..22C}.
 
We have developed a data-driven SFT model to capture solar surface magnetic field dynamics over the last century. For a description of this model see the Methods section. We extract the polar field information from this SFT model at every cycle minimum and utilize this as an input in a solar dynamo model to simulate the century-scale evolution of the sunspot-forming toroidal field component. The century-scale calibrated simulation, which is able to successfully reproduce solar activity over the past century is then used to predict the maximum (strength) of sunspot cycle 25 and its timing. Furthermore, we perform ensemble runs with expected level of fluctuations in the governing parameters of the solar cycle to generate a predicted range for cycle 25.  

\section*{Results}

{\bf Data-driven century-scale surface flux transport simulations.} Using the observed BMR emergence statistics we perform a continuous century-scale, data-driven simulation with our SFT model covering the period 1913-2016, i.e., from solar cycle 15 to the current cycle 24. The most important observed BMR statistics pertain to the flux, tilt angle, location and timing of the emergence of BMRs on the solar surface (details are available in the Methods section).

\begin{figure}[!h]
\centering
\includegraphics[height=10cm, width=18cm] {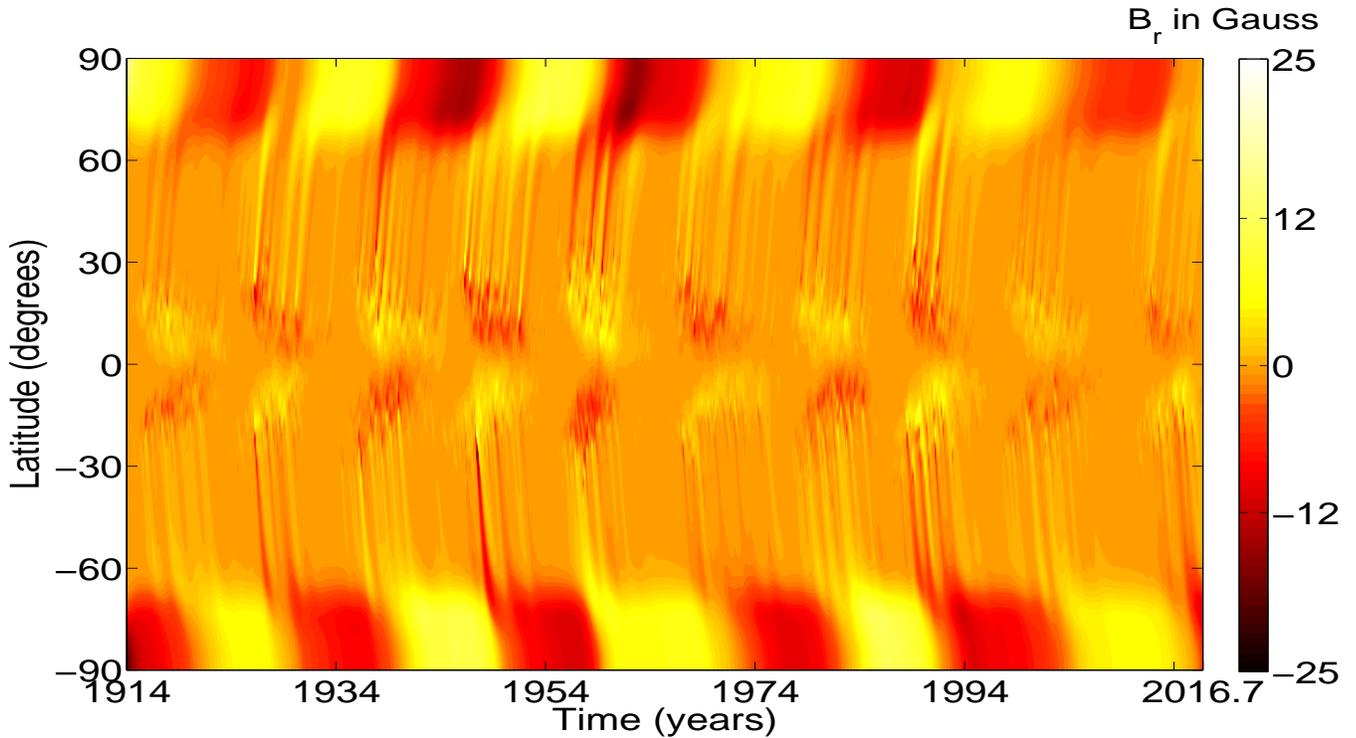}
\caption*{{\bf Fig. 1. Simulated butterfly diagram of the solar cycle.} Century-scale simulation of solar surface magnetic field evolution covering sunspot cycle 15 to the currently on-going cycle 24. The  butterfly diagram depicts the spatio-temporal variation of the longitudinally averaged radial magnetic field (in Gauss) on the Sun's surface.}
\label{fig1}
\end{figure}

In Fig. 1, we plot the longitudinally averaged radial magnetic field $B_r(R_{\odot},\lambda,\phi,t)$ as a function of latitude and time to generate the solar butterfly diagram corresponding to our simulation. Surface flux transport dynamics leading to solar polar field reversal and build-up is clearly evident. The top panel of Fig. 2a depicts the time evolution of total unsigned flux [$\Phi_k$, calculated by using equation (8) in Methods section] associated with the observed sunspot emergence statistics on the solar surface, which is used as input to drive the SFT simulation.

Quantitative comparison between our simulated polar field and observations is achieved through estimates of the simulated polar flux. We calculate the simulated polar flux ($\Phi^{N/S}_p$) by integrating the radial magnetic field around the polar cap (extending from $\pm$70$^{\circ}$ to $\pm $90$^{\circ}$) in both hemispheres [see equation (9) in Methods section]. During most of the period covered by the simulation, however, we do not have any data on how the magnetic field was spatially distributed -- precluding a direct estimate of the observed polar-cap flux. Observational magnetogram data is only available from 1975 onwards. Therefore, we rely on polar flux measurements obtained from MWO calibrated polar faculae data from 1906 onward\cite{2012ApJ...753..146M} for comparing our simulations to observation. The MWO polar flux data ends in 2014.5. Therefore we utilize polar field data provided by the Wilcox Solar Observatory (WSO) for the period beyond 2014.5.

The century scale simulation has to be initiated with a dipole moment. Given the unavailability of historical observation of dipole moment and the uncertainty in the polar faculae measurement, we vary the initial dipole moment strength by $\pm$25$\%$ and conduct multiple (continuous) 100 years runs. We select the simulation for which the correlation between the observed and simulated polar flux is maximum as our calibrated, standard simulation. Fig. 2b compares the time evolution of the simulated ($\Phi^{N/S}_p$) and observed polar flux for this calibrated simulation. As expected, the choice of the (arbitrary) initial magnetic field affects the polar flux generated early in the simulation (resulting in a disparity with observations at the end of cycle 15). Otherwise, the simulated polar flux is in good agreement with observations (within error bars) for most subsequent cycles. Linear correlation analysis between the simulated and observed polar flux amplitude of the two solar hemispheres at cycle minima gives a Pearson's correlation coefficient of 0.88 at a 99.99$\%$ confidence level. We note that over the past century, the only significant anomaly between observations and our simulation-based reconstruction is for the minimum of cycle 18. Exclusion of the northern and southern hemispheric polar flux values at cycle 18 minimum results in a Pearson's correlation coefficient of 0.95 (with 99.99$\%$ confidence level). This analysis is based on simulations up to September 2016 until which period observed sunspot emergence statistics were utilized to drive our simulations.

To forward-run the SFT model from the epoch when observed sunspot data inputs were stopped (i.e., September 2016, marked by the vertical dashed line in Fig. 2), synthetic input profiles are used to model the decaying phase of cycle 24 up to the end of 2019. We rely on various statistical properties of sunspots for modelling the synthetic profile to represent the plausible solar activity till the expected minimum (see Methods section for a detailed description). Once a synthetic profile is constructed, we simulate the last 3.25 years of cycle 24 using this input profile. The solid blue line in Fig. 2a represents a synthetic input profile that best fits the preceding phase of cycle 24. We use the calibrated simulation to forward run the model to predict the future evolution of the Sun's polar flux utilizing this synthetic input. We obtain a predicted polar flux value of 6.91 $\times$ 10$^{21}$ Maxwells in the northern hemisphere and $-$8.73 $\times$ 10$^{21}$ Maxwells in the southern hemisphere at the end of cycle 24.

A comparison of our predicted polar flux at cycle 24 minimum relative to previous cycles presented in Fig. 2b indicates that the south polar flux at the upcoming minimum of cycle 24 is likely to be stronger than the previous minimum while the north polar flux may not be significantly different from the previous minimum.

{\bf Ensemble forecast of the Sun's polar field.} Towards generating an ensemble forecast to ascertain the range (uncertainty) around our standard run based polar field prediction we first simulate the decaying phase of cycle 24 with thirty-three other realizations of the sunspot input profile (represented by the set of green curves in Fig. 2a). Among these synthetic profiles, 24 profiles are constructed based on varying the amount of total flux associated with the cycle by $\pm$30$\%$ around the mean. Five additional synthetic profiles are constructed by changing the latitudinal width of the activity wings. The remaining four profiles are obtained by redistributing the latitudinal position of the same sunspots (keeping their time of emergence unaffected). We find that the maximum spread in the predicted value of polar flux is obtained when the total flux in the synthetic profiles is varied. The variation in the flux of the synthetic input results in the predicted northern hemispheric polar flux (during cycle 24 minimum) varying within the range (6.13 -- 7.29) $\times$ 10$^{21}$ Maxwells (cyan lines beyond September 2016 in Fig. 2b); the southern hemispheric polar flux varies within (7.91 -- 9.41) $\times$ 10$^{21}$ Maxwells (magenta colored lines in Fig. 2b). Variation in latitudinal spread results in a smaller spread in polar flux value. The third class of profiles (redistribution of latitudinal position of sunspots) do not produce substantial variations in polar flux amplitude.

\begin{figure}[!ht]
\centering
\includegraphics[height=15.2cm, width=18 cm]{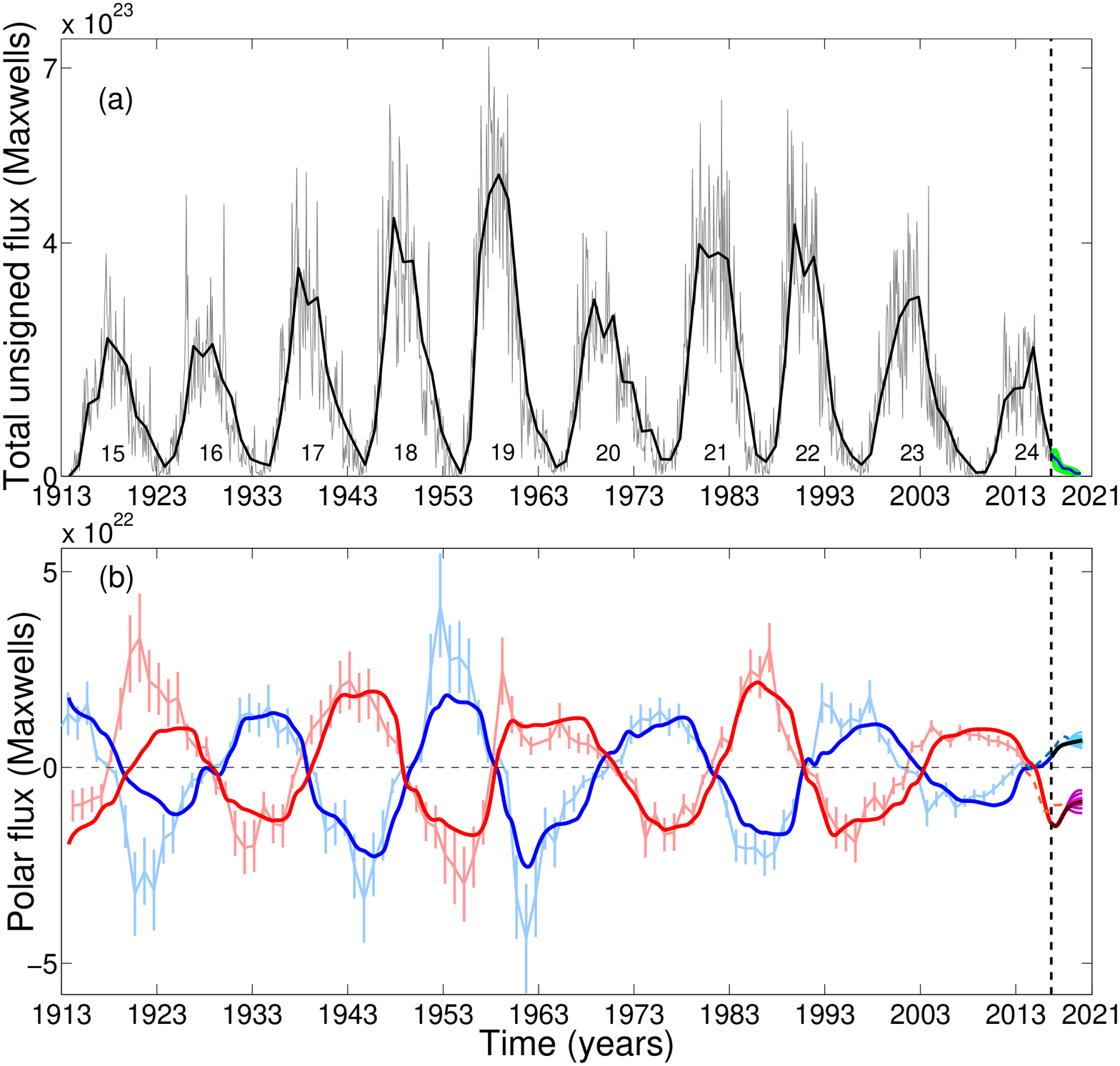}
\caption*{{\bf Fig. 2. Data-driven simulation of solar surface polar flux.} (a) This panel depicts the evolution of total unsigned flux associated with sunspots which is used to drive the surface flux transport model until September 2016 (epoch marked by the dashed vertical line). The gray and solid black curves represent monthly and annually averaged unsigned flux, respectively. The blue and set of green curves beyond September 2016 depicts different realizations of the modelled descending phase of cycle 24 used as synthetic inputs to forward run the surface flux transport model. (b) Time evolution of polar flux calculated from our surface flux transport simulation compared with those obtained from polar faculae observations. The light blue (and light red) curve with error bars represents the polar flux estimated from polar faculae observations in the northern (and the southern) hemisphere, whereas the solid blue (and red) curve shows the polar flux obtained from our simulation for corresponding hemispheres. The dashed blue (and dashed red) curve beyond 2014.5 represents the polar flux obtained from WSO polar field observations for the northern (and the southern) hemisphere. The set of light and dark cyan and magenta curves beyond September 2016 (vertical dashed line) depict predicted polar fluxes (up to 2020) from ensemble runs with varying input fluxes and tilt angle fluctuations, respectively. The black and dark red curves represent the polar field prediction from our standard run.}
\label{fig2}
\end{figure}

We note that deviations between the simulated and observed polar flux may arise due to assumptions that are necessary in order to perform such long-term simulations, especially where observational constraints are limited. We have assumed all sunspots appearing on the solar surface follow mean statistical properties and are ideal BMRs (which fall under $\beta$-type configuration of sunspots). However, other types of magnetic configurations are possible: \textit{$\alpha$}, \textit{$\gamma$}, \textit{$\delta$} or a combination of these. We also do not incorporate any scatter in the tilt angle distribution of sunspots in our model. Detailed studies of BMR tilt angle distribution\cite{2010A&A...518A...7D,1999SoPh..189...69S,2013SoPh..287..215M} have established Joy's law, but also found a scatter of individual tilt angles about the mean. Studies of the effect of tilt angle scatter on the polar field and dipole moment evolution\cite{2014ApJ...791....5J,2017SoPh..292..167N} has established that large sunspots with a large scatter in tilt angle leave a notable imprint on polar field amplitude. Furthermore, a large individual sunspot with an orientation that is opposite to what is expected in a particular hemisphere for a particular cycle (i.e., a non-Hale region) appearing at lower latitudes can cause a significant decrease in polar field strength\cite{2015SoPh..290.3189Y}. Fluctuations in the meridional circulation may also impact polar field amplitude\cite{2014ApJ...792..142U}.

To further examine the impact of these irregularities in the sunspot cycle on our prediction range we consider some probable scenarios in keeping with the philosophy of our ensemble forecast. To explore the effect of the occurrence of non-Hale regions in the last 3.25 years of cycle 24, we introduce 10--20 non-Hale BMRs, which constitute about 3--6$\%$ of the total flux associated with the input profile. This results in a 3--5$\%$ decrease in the final polar flux value calculated during cycle 24 minimum. Introducing $\pm$30$\%$ fluctuations in the peak speed of the meridional circulation during the last 3.25 years of cycle 24 results in $\pm$6.5$\%$ (on average) variation in the polar field generated at cycle minimum. We further explore the impact of introducing randomness in the tilt angle distribution of active regions. We perform 110 additional simulations with plausible tilt angle fluctuations incorporated in the input profile for the descending phase of cycle 24. The tilt angle fluctuations are constrained by solar cycle observations (see the Methods section for a detailed description). Incorporating tilt angle scatter further increases the predicted polar flux range at cycle 24 minimum to $\pm$30$\%$, on average, in both the hemispheres (as depicted in Fig. 2b). We note, therefore, that the range of variation induced by incorporating tilt angle fluctuations exceed (and subsume) variations due to other processes. This extensive parameter dependence study indicates the robustness of our simulations, and the ensemble forecast provides a predicted range of solar polar field at the end of cycle 24 minimum. 

\noindent {\bf Prediction of solar cycle 25 using a dynamo model}. We obtain a strong correlation between the simulated polar flux (averaged over two hemispheres) at cycle minimum and the amplitude of the next cycle. We obtain a Pearson's linear correlation coefficient of 0.84 with a confidence level of 99.10$\%$. This reiterates that the previous cycle polar flux is the best proxy for predicting the sunspot cycle. The connection between these two quantities can be explained by solar dynamo theory. The polar field originates from the poloidal component of the Sun's magnetic field whereas the sunspot cycle amplitude is governed by the strength of the toroidal component of the magnetic field. The latter, however, is generated by stretching of the poloidal component by differential rotation. Thus, the initial poloidal field seed of cycle ($n-1$, say) (of which the polar field is a proxy) directly governs the strength of the toroidal field of cycle ($n$) -- which in turn governs the strength of the associated sunspot cycle ($n$). Therefore, we utilize the simulated polar field from the SFT model and its corresponding poloidal field at cycle minima in a dynamo model of the solar interior to predict the amplitude of the next cycle toroidal field and hence the strength of the associated sunspot cycle.

\begin{figure}[!h]
\centering
\includegraphics[height=9cm, width=18cm]{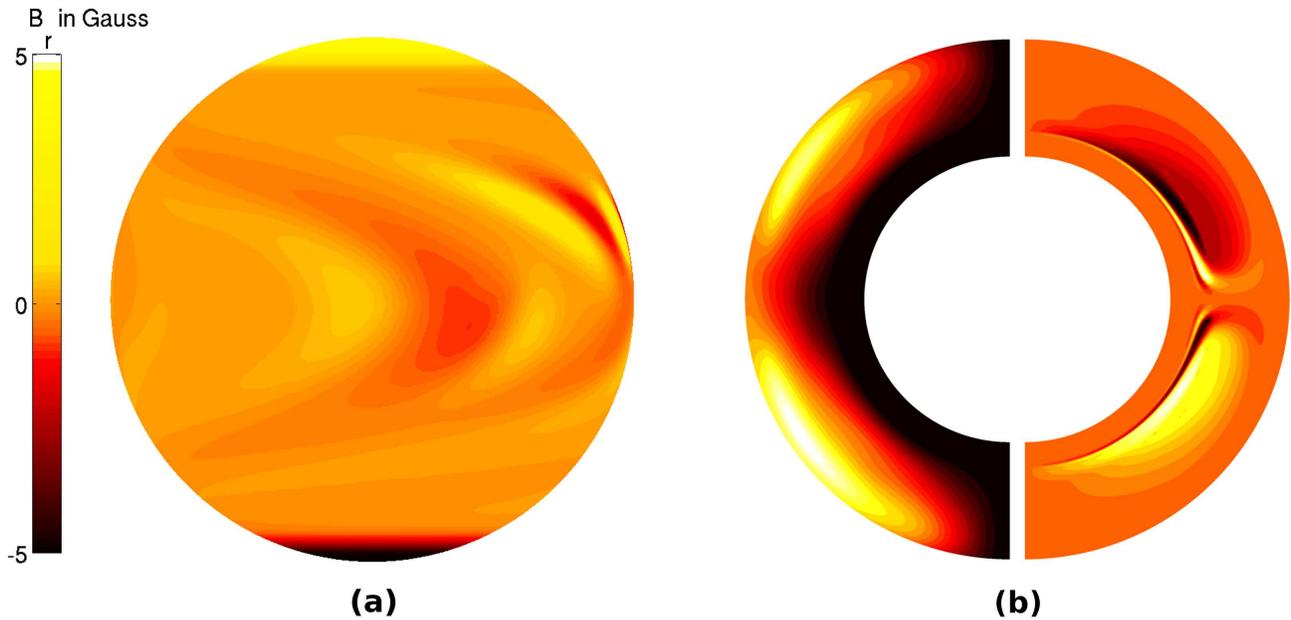}
\caption*{{\bf Fig. 3. Dynamo simulations of the Sun's internal magnetic field driven by surface inputs}. (a) Solar surface flux transport model prediction of the Sun's surface magnetic field distribution near cycle 24 minimum. (b) {\bf Left:} The poloidal field distribution within the Sun's convection zone in the internal dynamo model following assimilation of the data from the surface flux transport model. (b) {\bf Right:} The dynamo model predicted toroidal field within the Sun's convection zone during cycle 25 maximum.}
\label{fig3}
\end{figure}

From our data-driven SFT simulation we first calculate the surface magnetic field (averaged over $\phi$) during the minimum of each solar cycle starting from cycle 16 minimum (1934). This surface magnetic field map is assimilated (at the corresponding cycle minima) in simulations with a kinematic, axisymmetric dynamo model\cite{2014A&A...563A..18P} (following calibration and processing as detailed in the Methods section). This century-scale dynamo simulation is forward run in a predictive mode to simulate solar cycle 25 with regular ``correction'' of the poloidal component at every cycle minimum. We note that our method of utilizing the SFT output as an input in a dynamo model only at cycle minima is distinct from a fully coupled SFT-dynamo simulation\cite{2017ApJ...834..133L} and herein, the methodology is devised for predictive purposes utilizing the understanding that has been recently established\cite{2008ApJ...673..544Y,2010LRSP....7....6P,2012ApJ...761L..13K}. In Fig. 3a, an output from our SFT simulation is presented which depicts the predicted surface distribution of magnetic field during cycle 24 minimum. The left-hand panel in Fig. 3b represents the magnetic vector potential (corresponding to the poloidal field) in the dynamo model 20 days after the SFT derived vector potential is assimilated into the dynamo model. Using this vector potential as an input in the dynamo model, and running this forward in time, we generate the predicted shape, strength and timing of sunspot cycle 25. Fig. 3b represents the distribution of the toroidal magnetic field in the Sun's convection zone at the maximum of sunspot cycle 25.

This multi-cycle continuous solar dynamo simulation with assimilation of poloidal field maps from the data-driven surface flux transport simulation (at cycle minima) is used to simulate the sunspot cycle over century scale. The cycle strength is determined by the dynamo simulated toroidal flux eruption based on an in-built buoyancy algorithm which models sunspot eruptions\cite{2014A&A...563A..18P,2002Sci...296.1671N,2004A&A...427.1019C}. This is compared to sunspot cycle observations in Fig. 4 (following an appropriate multi-cycle calibration as detailed in the Methods section). We recover a good correlation between the yearly averaged simulated and observed sunspot cycle amplitudes. We obtain a Pearson's linear correlation coefficient of 0.87 with a confidence level of 99.54$\%$; exclusion of cycle 19 from the correlation analysis (whose deviation from observations is a result of the mismatch between simulated and observed polar field at cycle 18 minimum) generates an improved correlation of 0.98 with 99.99$\%$ confidence level. The significant agreement between dynamo simulations in the predictive-mode and past sunspot cycle observations lays the foundations of our prediction of the toroidal component of solar cycle 25 based on our dynamo model.

\begin{figure}[!h]
\centering
\includegraphics[height=9.3cm, width=17cm] {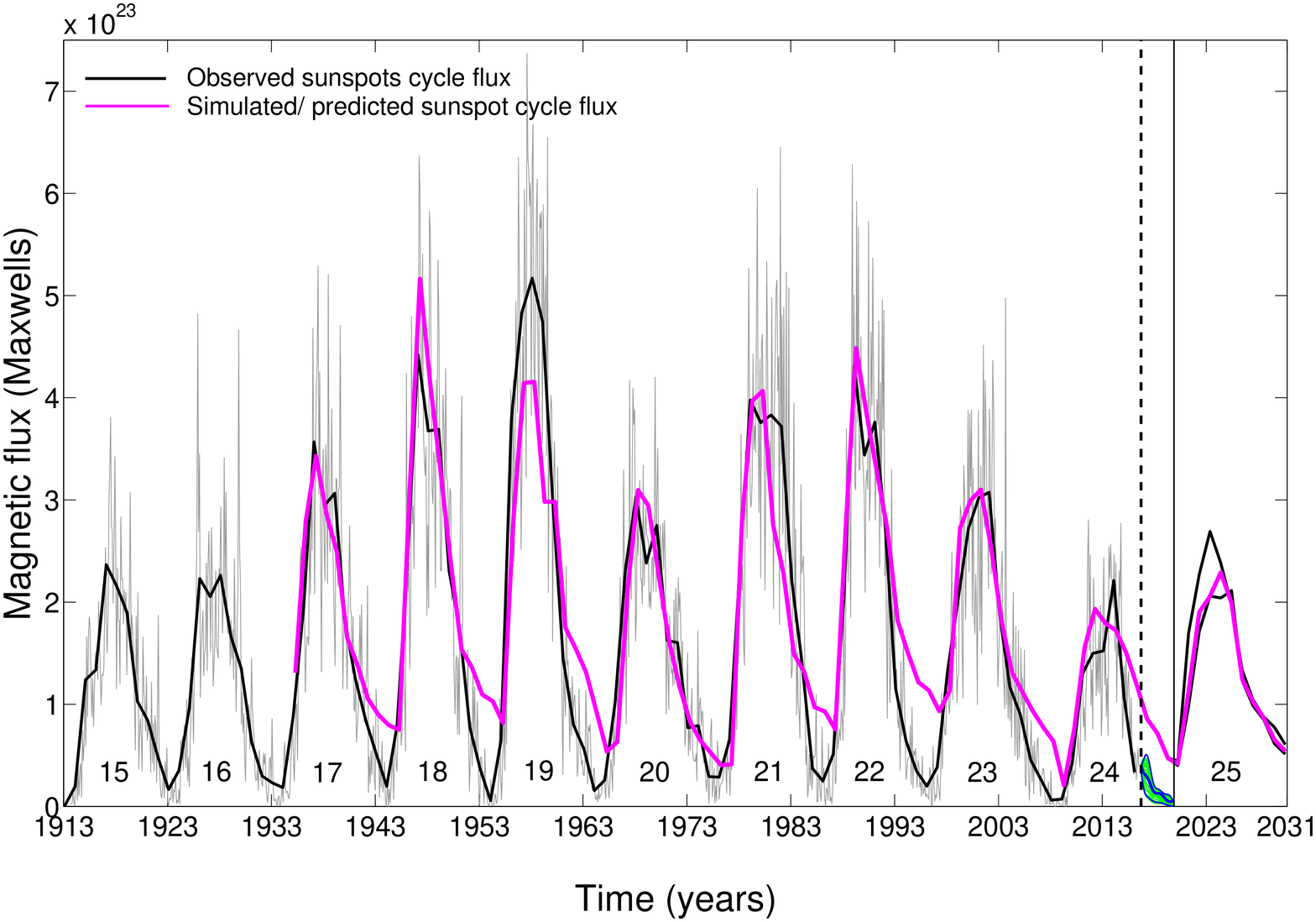}
\caption*{{\bf Fig. 4. Prediction of sunspot cycle 25.} Solar dynamo simulated sunspot cycles (magenta curve) compared with the observed sunspot cycle (unsigned magnetic flux; black curve), where both quantities are yearly averaged. The light gray curve in the background represents monthly averaged unsigned sunspot flux. The blue and set of green curves between the vertical black dashed and solid lines depict flux associated with the thirty-four synthetic profiles used in the surface flux transport model as plausible realizations of the descending phase of cycle 24. The magenta curve beyond the solid-black vertical line (corresponding to our standard simulation) depicts the predicted shape and strength of sunspot cycle 25. The set of two black curves beyond the solid-black vertical line represent the strongest and the weakest magnetic cycles (that is the range of our ensemble forecast) based on our diverse predictive dynamo runs. The prediction range (uncertainty) indicates cycle 25 will be similar or slightly stronger than the previous cycle.}
\label{fig4}
\end{figure}

Our prediction of cycle 25 is presented in Fig. 4. We accommodate the range of possibilities (uncertainty) from our ensemble forecast for the solar polar field at the minimum of cycle 24 in the following way. In addition to the standard predictive-mode surface flux transport run, we select the runs corresponding to the strongest and weakest polar flux realizations (at cycle 24 minima) from our set of ensemble forecast runs to drive three distinct predictive mode dynamo runs. The magenta curve beyond 2020 (marked by a vertical black solid line) in Fig. 4 depicts the dynamo predicted evolution of the buoyantly emerging magnetic flux for the standard run, which generates our most likely prediction. The two black curves beyond 2020 depict the range of our prediction based on the extreme realizations from our ensemble forecast which is found to be (2.11--2.69)$\times 10^{23}$ Maxwells. Our prediction (Fig. 4) shows that sunspot cycle 25 would be similar or slightly stronger in strength relative to the current cycle 24 with the standard simulation-run (magenta curve) generating a peak amplitude of 2.29$\times 10^{23}$ Maxwells. A calibration with the observed amplitude of the annually averaged sunspot number time series also yields a prediction for the strength of cycle 25 in terms of the sunspot number -- which is used more often in statistical or empirical forecasts. Based on our simulations, the corresponding prediction for the yearly mean sunspot number at the maximum of solar cycle 25 is 118 with a predicted range of 109--139.

\vspace{2cm}
\begin{table}[!h]
\centering
\begin{tabular}{ |p{4.5cm}|p{2.0cm}|l|l|l|l| }
\hline
\multicolumn{3}{ |c| }{{ \bf Table 1}: Amplitude and Timing of the Maximum of Solar Cycle 25} \\
\hline
{} & Prediction & Range \\ 
\hline
&  &  \\ 
Flux (Maxwells) & 2.29 $\times$ 10$^{23}$ & (2.69 -- 2.11)$\times$ 10$^{23}$  \\ 
&  &  \\ 
\hline 
&  &  \\ 
Yearly mean sunspot number & 118 & 139 -- 109 \\
&  &  \\ 
\hline
&  &  \\ 
Time of peak & 2024 & 2023 -- 2025 \\
&  &  \\ 
\hline
\end{tabular}
\end{table}

Although, we did not make any specific attempts to do so, serendipitously, we find that the timing of the peak of simulated sunspot cycles matches quite well the timing of the maxima of observed sunspot cycles to within half a year on average (Fig. 4). We believe this is due to the ability of our dynamo model to match the cycle amplitudes and an existing (observed) empirical relationship between the amplitude and the rise-time of sunspot cycles known as the Waldmeier effect\cite{2002SoPh..211..357H}. The dynamo model self-consistently imbibes this effect resulting in the close correspondence of simulated and observed cycle-peak timings. Taking advantage of this, and relying upon our dynamo forward run for solar cycle 25, we infer that the maximum of solar cycle 25 will occur around 2024($\pm$1). The range of $\pm$1 year also includes the uncertainty in the exact timing of cycle 24 minimum which may vary by six months.

In Table 1 we summarize the predicted properties of sunspot cycle 25, including its amplitude, timing and range (uncertainty) derived from our ensemble forecast. 

\section*{Discussion}

In summary, we have utilized a solar surface flux transport model and a solar internal dynamo model for the first, continuous century-scale calibrated simulation of solar activity. We emphasize that no SFT or dynamo model parameters were tuned after the simulations were initialized, and the simulated variations are thus a direct result of data assimilation from the SFT model to the dynamo model, only. Except cycle 19 -- the strongest and most extreme cycle in the last century -- our simulations reproduce past solar activity peaks (and their relative variations) quite well. We note that the ``floor'' in the dynamo simulated toroidal field activity during minimum phases does not reach the observed lows; however, this has no impact on the peak cycle strengths or their timing -- which is our focus here. 

We emphasize that the excellent agreement between surface flux transport enabled simulations of solar activity and the strength of sunspot cycles over the past century, lends strong, independent support to the emerging view that the Babcock-Leighton mechanism\cite{1961ApJ...133..572B,1969ApJ...156....1L} is currently the dominant solar poloidal field creation mechanism in the Sun\cite{2010A&A...518A...7D,2015Sci...347.1333C,2013ApJ...767L..25M}. Our simulations indicate that no other mechanism of variability is necessary to explain the observed variability in the solar cycle over the last 100 years. However, we do find that it is important to maintain a (non-varying) source or seed of weak magnetic field at all times (see Methods section), which has been found necessary in independent studies \cite{2014A&A...563A..18P,2014ApJ...789....5H} and indeed expected in the solar convection zone\cite{2010LRSP....7....3C}. In this light, our simulations and its agreement with century-scale solar activity observations strongly suggests that while a source of weak magnetic fields in the solar interior (such as the mean-field dynamo $\alpha$-effect) may be necessary, the variability of the amplitude of sunspot cycles over decadal to century-scale is primarily governed by the variability in the tilt-angle and flux distribution of bipolar sunspot pairs or BMRs and their surface evolution. 

We demonstrate that the prediction window for solar cycles can be extended to a decade allowing for advanced space weather assessment and preparedness. We achieve this by assimilating data from a solar surface magnetic field evolution model to a solar internal dynamo model run in the predictive mode. We predict a weak, but nonetheless, not insignificant solar cycle 25. Our ensemble forecast involving about 140 plausible realizations of the solar surface polar field provides a prediction range that indicates that sunspot cycle 25 would be similar or somewhat stronger than the current cycle 24. Based on our simulations we additionally predict that the maximum of sunspot cycle 25 will occur around 2024($\pm$1). 

It is important to note here the substantial progress achieved over the last decade in our understanding of the predictability of sunspot cycles\cite{2008ApJ...673..544Y,2012ApJ...761L..13K,2013ApJ...767L..25M} -- which was spurred by substantial disagreements and controversy surrounding predictions for sunspot cycle 24\cite{2008SoPh..252..209P}. This emergent understanding has seeded diverse physics-based predictive approaches increasing in complexity -- assessments based on solar surface flux transport models\cite{2016JGRA..12110744H,2016ApJ...823L..22C}, semi-empirical forecasts combining surface flux transport models with solar cycle statistics\cite{2018ApJ...863..159J} -- whose results seem to be convergent with our more sophisticated century-scale data assimilation approach coupling a magnetic flux transport model on the surface to a magnetohydrodynamics dynamo model in the Sun's interior. 

Our ensemble prediction indicates the possibility of a somewhat stronger cycle than hitherto expected, which is likely to buck the significant multi-cycle weakening trend in solar activity. Our results certainly rule out a substantially weaker cycle 25 compared to cycle 24 and therefore, do not support mounting expectations of an imminent slide to a Maunder-like grand minimum in solar activity. This had given rise to associated speculations regarding a period of global cooling (in the Earth's climate); these findings negate such possibilities at least over the next decade or so.

We conclude that near-Earth and inter-planetary space environmental conditions and solar radiative forcing of climate over sunspot cycle 25 (i.e., the next decade) will likely be similar or marginally more extreme relative to what has been observed during the past decade over the current solar cycle.

\section*{Methods}
\subsection*{The Surface Flux Transport (SFT) Model}

\noindent {\bf The basic equation.} We have developed a new model to study the evolution of the Sun's photospheric magnetic field ($\mathbf{B}$) which is governed by the magnetic induction equation,

\begin{equation}
\frac{\partial \mathbf{B}}{\partial t} ={\nabla} \times (\mathbf{v} \times \mathbf{B}) + \eta \nabla^{2} \mathbf{B}
\end{equation}

\noindent Where $\mathbf{v}$ represents the large scale velocities (both meridional circulation and differential rotation) responsible for advection of $\mathbf{B}$ and $\eta$ represents the magnetic diffusion. Since most of the surface magnetic field is confined in the radial direction\cite{1993SSRv...63....1S}, we shall solve only for the radial component of the field. The radial component $B_r(\theta, \phi,t)$ of the induction equation when expressed in spherical polar coordinates is,

\begin{multline}
\frac{\partial B_r}{\partial t} = -  \omega(\theta)\frac{\partial B_r}{\partial \phi} - \frac{1}{R_\odot \sin \theta} \frac{\partial}{\partial \theta}\bigg(v(\theta)B_r \sin \theta \bigg)
+\frac{\eta_h}{R_\odot^2}\bigg[\frac{1}{\sin \theta} \frac{\partial}{\partial \theta}\bigg(\sin \theta \frac{\partial B_r}{\partial \theta}\bigg) + \frac{1}{\sin \theta ^2}\frac{\partial ^ 2 B_r}{\partial \phi ^2}\bigg] + S(\theta, \phi, t)
\end{multline}

\noindent Here $\theta$ is the co-latitude, $\phi$ is the longitude, $R_\odot$ is the solar radius, $\omega(\theta)$ is the differential rotation and $v(\theta)$ is the meridional circulation on the solar surface. The parameter $\eta_h$ is the effective diffusion coefficient and $ S(\theta, \phi, t)$ is the source term describing the emergence of new sunspots. Since we are studying the evolution of $B_r$ on the surface of a sphere, the code has been developed using spherical harmonics.

\noindent {\bf Input parameters.} The differential rotation is a large scale plasma flow along $\phi$ and it varies with latitude. Therefore, the latitudinal shear of differential rotation stretches the magnetic field along the $\phi$ direction. As the leading and trailing spots of a tilted BMR reside in different latitudes, the differential rotation increases the longitudinal distance between the two polarities of the same BMR. The differential rotation has been modelled using an empirical profile\cite{1983ApJ...270..288S}.

\begin{equation}
\omega(\theta) = 13.38 - 2.30 \cos^{2}\theta - 1.62 \cos^{4}\theta
\end{equation}

\noindent Wherein, $\omega(\theta)$ has units in degrees per day. This profile has been validated by recent helioseismic observations\cite{1998ApJ...505..390S}. The supergranular cells in the solar convection zone effectively diffuse the magnetic field on the solar surface. Some models\cite{2001ApJ...547..475S,1994ApJ...430..399W} have treated convective motion of supergranules as a discrete random walk process, rather than using a fixed diffusion coefficient, while others\cite{2016JGRA..12110744H,2014ApJ...792..142U}-s, have considered a purely advective flux transport model where the convective flows of supergranules are included as a part of the velocity profile. However, we have taken a constant value of the diffusion coefficient ($\eta_h$ as 250 km${^2}$s$^{-1}$) which lies within the range of observed values\cite{2000ssma.book.....S}. Another large-scale flow, i.e., the meridional circulation on the solar surface carries magnetic field from lower latitudes to higher latitudes. Though the flow profiles and peak amplitude of meridional circulation vary from model to model\cite{2012LRSP....9....6M}, they all have some fundamental similarities. The flow speed becomes zero at the equator and the poles, and the circulation attains its peak velocity (10 - 20 ms$^{-1}$) near mid-latitude. To replicate this large-scale flow we have used a velocity profile prescribed by van Ballegooijen\cite{1998ApJ...501..866V},

\begin{equation}
 v(\lambda) =
  \begin{cases}
    - v_{0} \hspace{0.1cm \sin( \pi \lambda/ \lambda_0)} & \text{if } |\lambda | < \lambda_0  \\
   0       & \text{otherwise }
  \end{cases}
\end{equation}

\noindent where $\lambda$ is the latitude in degrees ($\lambda = \pi/2 - \theta$) and $\lambda_0$ is the latitude beyond which the circulation speed becomes zero. In our model we have taken $\lambda_0 = 75^o$ and $v_0$ =15 ms$^{-1}$.

Simulations for multiple solar cycles using observed cycle amplitudes show that the polar field systematically drifts and eventually fails to reverse its sign. There are three prescribed ways to address this problem: 1) varying the meridional circulation amplitude from cycle to cycle according to cycle strength\cite{2002ApJ...577L..53W}; 2) including an additional radial diffusivity which forces polar field to decay at a time scale of 5 years\cite{2002ApJ...577.1006S,2006A&A...446..307B} and 3) introducing a modified Joy's law, in which the tilt angles of BMRs depend on both latitude and cycle strength\cite{2010ApJ...719..264C}. The later is physically motivated and supported by independent simulations of the buoyant rise of flux tubes and we have followed this third prescription in our SFT model.

\noindent {\bf Replication of flux emergence of sunspots.} Modelling of flux emergence requires information of the position of sunspots on the solar surface and the area associated with the spots. Since we do not model the growth of sunspots in our simulation, we take data at the time of their maximum surface area rather than their time of appearance on the photosphere. We assume all sunspots that appear on the photosphere are BMRs (i.e., type $\beta$). Location, area and time information of sunspots are provided by the Royal Greenwich Observatory (RGO) and USAF/NOAA database. Beyond year 1976, the source of information associated with the active regions changes from RGO to USAF/NOAA. To maintain the consistency in area measurement from two different data sources, we multiply a constant factor of 1.4 to any active region area belonging to USAF/NOAA database if its area is smaller than 206 micro-hemispheres which corresponds to a pair of sunspots each with radius of 10 Megameters -- such cross-database calibration is often necessary due to different instrument specifications and diverse record-keeping practices \cite{2015ApJ...800...48M,2014SoPh..289.1517F}. Multiple consistency checks that we performed independently point out the necessity of this correction for consistency between the RGO and USAF/NOAA sunspot area databases. 

We calculate the flux associated with a BMR using an empirical relationship\cite{2006GeoRL..33.5102D}: $\Phi(A) = 7.0 \times 10 ^{19} A$ Maxwells, where $A$ is the area of the whole sunspot in units of micro-hemispheres. This flux is equally distributed among the two polarities of the BMR. Also, we can easily determine the value of radius (say, R${_{\text{spot}}}$) for each of the leading and following polarities from the area information. We assume that the radial separation (say, $d$) between the centroids of leading and following spots is proportional to R${_{\text{spot}}}$. The tilt angle ($\alpha$) of the BMR is assigned in the following manner,

\begin{equation}
\alpha = g T_{n} \sqrt{|{\lambda}|}
\label{tilt}
\end{equation}

\noindent where $\lambda$ is the latitudinal position of the centroid of the whole BMR. The quantity $T_n$ accounts for the variation of tilt angle with cycle strength\cite{2011A&A...528A..82J}. The factor g is introduced to include the effect of localized inflows towards active regions, that is present on the photosphere apart from the large-scale inflows related to activity belts. These localized inflows effectively reduce the latitudinal separation between opposite polarities and allow less flux to reach the polar region.

Since the polar flux is proportional to tilt angle, we incorporate the impact of these localized inflows by reducing the tilt angle\cite{2010ApJ...719..264C}. We choose g to be equal to 0.7. Once we know the location of the centroid of the whole BMR in co-latitude($\theta^c$) and longitude($\phi^c$), the positions of individual polarities of the BMR are decided as follows: $\theta^{l/f} = \theta^c \pm \frac{d}{2}\sin \alpha$ and $\phi^{l/f} = \phi^c \pm \frac{d}{2}  \cos \alpha$, where '$l$' and '$f$' denote the leading and the following polarities respectively. The initial radial magnetic field associated with the BMR is

\begin{equation}
\Delta \text{B}_r(R_{\odot},\theta,\phi) = \text{B}^l_r(R_{\odot},\theta,\phi) - \text{B}^f_r(R_{\odot},\theta,\phi)
\end{equation}

\noindent where $\text{B}^{l/f}_r (R_{\odot},\theta,\phi)$ are the unsigned magnetic field distribution of the leading and following spots which have opposite polarities. Each spot is modelled as\cite{1998ApJ...501..866V},

\begin{equation}
\text{B}^{l/f}_r (R_{\odot},\theta,\phi) = \text{B}_{\text{max}} \text{exp} \Big\{- \frac{2[1-\cos \beta^{l/f}(\theta,\phi)]}{0.16 R_{\text{spot}}^2}\Big\}
\end{equation}

\noindent where $\beta^{l/f}$ are the heliographic angle between ($\theta, \phi$) and the central coordinates of the leading and following polarities ($\theta^{l/f},\phi^{l/f}$) respectively. $\text{B}_{\text{max}}$ is the maximum value of magnetic field of each polarity, which is automatically decided by the flux contained in the spot.\\

\noindent {\bf Initial field configuration.} We use our SFT model to study the evolution of the large-scale photospheric magnetic field for multiple solar cycles, starting from solar cycle 15 around the year 1913. As we do not have any full-sun magnetic field data at the beginning of cycle 15, we use an axisymmetric dipolar configuration\cite{1998ApJ...501..866V} as an initial field condition to initiate our simulations. We have tried to minimize the difference between the polar flux associated with this initial field and the polar flux at the beginning of cycle 15 acquired from polar faculae observations\cite{2012ApJ...753..146M}. However, the actual magnetic field configuration at the beginning of cycle 15 may substantially differ from our choice of initial field. This arbitrariness leads us to exclude the polar field produced by our simulation at the end of cycle 15 from any correlation study or calibration of our model.\\

\noindent {\bf Numerical modeling parameters.} Ideally one should consider all possible values of degree ($l$) of spherical harmonics. Instead of taking the full range of values of $l$ from 0 to $\infty$, we consider $l$ values varying from 0 to 63. Our choice is motivated by the fact that $l = 63$ corresponds to a typical size of the supergranular cells (roughly 30 Mm) on the solar photosphere.

\subsection*{Measured Quantities}

The quantity plotted in Fig. 2a is the total unsigned magnetic flux associated with the sunspots emerging during the period spanning from the year 1913 to 2031. It includes data from direct sunspot observations (depicted by the gray curve) and also the constructed decaying phase of cycle 24 (blue and green curves). If we assume n${_k}$ is the total number of individual spots appearing on the solar surface in the k${^{th}}$ month and area of those individual spots are A${_i}$ ({\it i} = 1,2,...,n${_k}$); the total unsigned flux associated with the emerging spots ($\Phi_{k}$, denoted by the gray curve in Fig. 2a) in the k${^{th}}$ month would be
\begin{equation}
\Phi_{k} = \sum_{i=1}^{n_{k}} \Phi_i(A_i)
\end{equation}
\noindent Wherein, $\Phi_i(A_i) = 7.0 \times 10 ^{19} A_i$ Maxwells.
We calculate the polar flux (plotted in Fig. 2b) by integrating radial magnetic field around the polar cap region (extending from $\pm$70$^{\circ}$ to $\pm$90$^{\circ}$) in both hemispheres and using the following equation

\begin{equation}
\Phi_p^{N/S} (t) = \int_{0^{\circ}}^{360^{\circ}}\int_{\pm {70^{\circ}}}^{\pm {90^{\circ}}}B_r(R_{\odot},\lambda,\phi,t) \cos\lambda d\lambda d\phi
\label{polarNS}
\end{equation}

\noindent where $\lambda$ is latitude and $\phi$ is longitude.

\subsection*{Construction of the synthetic sunspot input profiles}

We consider the magnetic flux to be a better proxy of solar activity than the sunspot numbers. Thus, our synthetic sunspot data profile is mainly based on the flux evolution observed in cycle 24 so far. The time evolution profile of sunspot number during a certain solar cycle can be determined by using a generalized function with the knowledge of its starting time and peak activity\cite{1994SoPh..151..177H}. For constructing a synthetic input profile, the observed sunspot data of cycle 24 (spanning over 2008.5--2016.75) is fitted with a mathematical function, an extension of which also models the remaining 3.25 years of the descending phase of cycle 24. We further introduce random fluctuations to this mean profile to produce a more realistic (observationally) input profile. The total number of sunspots associated with a typical synthetic profile is roughly 2800.

While assigning area to the spots associated with a synthetic profile, we follow a similar statistical distribution of area obtained by analyzing the observed sunspot data of cycle 24. For the time-latitude allocation of the emerging BMRs on the solar surface, we use an empirical functional form to calculate the mean latitude and the spread of the activity belts\cite{2011A&A...528A..82J}. The spots are randomly distributed over all possible longitude on the solar surface. The tilt angles of the BMRs are decided by equation (5). We get the flux associated with BMRs using the same relationship, $\Phi(A) = 7.0 \times 10 ^{19} A$ Maxwells. We constructed a set of thirty-four different synthetic input profiles by modulating total flux associated with the sunspots, or by varying the latitudinal spread (and interchanging their relative position) in the activity wings. Among these thirty-four profiles one closely follows the already observed (up to September 2016) sunspot distribution of cycle 24, and we regard this profile as a standard one. The modeling of twenty-four synthetic input profiles was done by varying the total sunspot-associated flux by $\pm$30$\%$ about the standard input profile. Once we have all particulars related to the sunspots of a certain synthetic profile, we consider only the last 3.25 years of the profile and add them to the existing observed sunspot data of cycle 24 to model likely emergence profile up to the end of 2019.

\noindent \textbf{Introduction of randomness in tilt angle of active region.}
We generate 110 synthetic input profiles where we introduce randomness in the tilt angles of active regions in addition to the systematic tilt which is entirely determined by Joy's law. The scatter around the systematic mean tilt angle decreases with increasing active region area such that the standard deviation of the distribution of tilt angle randomness follows a linear logarithmic relation with active region area\cite{2014ApJ...791....5J}. The tilt angle of each active region is determined by the relation, $\alpha = g T_{n} \sqrt{|{\lambda}|}+\epsilon$, where the first part in the right hand side is the same as equation (5) and the second part, $\epsilon$, represents the randomness\cite{2014ApJ...791....5J}. Every individual value of $\epsilon$ is selected randomly from a normal distribution with zero mean and a standard deviation that is decided by the area of the particular active region in consideration. We apply this method to every active region of the standard input profile and generate a set of 50 such realizations. We also consider 60 different input profiles where scatter in the tilt angle is introduced in the strongest and the weakest (according to total sunspot-associated flux) profiles to model the maximum uncertainty that can be present in the descending phase of cycle 24.

\subsection*{The Solar Dynamo Model}

Most of the existing solar dynamo models\cite{2008ApJ...673..544Y,2002Sci...296.1671N,2004A&A...427.1019C,1999ApJ...518..508D} identify Babcock-Leighton mechanism as the sole process for generation of the poloidal component (B$_{P}$) from the toroidal component (B$_{T}$) of the magnetic field. However, other studies\cite{2014A&A...563A..18P,2014ApJ...789....5H} have established the mean field $\alpha$-effect, present in the bulk of the SCZ, as an essential means for reproducing important observational features. In our case, we have used a solar dynamo model which includes both B-L mechanism and mean field $\alpha$-effect for conversion of B$_{P}$ from B$_{T}$. In the following section, we shortly describe the model that we have used. The same model has provided satisfactory results previously\cite{2014A&A...563A..18P}. The axisymmetric dynamo equations solved in kinematic regime are,

\begin{equation}
    \frac{\partial A}{\partial t} + \frac{1}{s}\left( \textbf{v}_p \cdot \nabla\right)  (sA) = \eta_p\left( \nabla^2 - \frac{1}{s^2}  \right)A + \alpha B,
\end{equation}

\begin{equation}
    \frac{\partial B}{\partial t}  + s\left[ \textbf{v}_p \cdot \nabla\left(\frac{B}{s} \right) \right]
    + (\nabla \cdot \textbf{v}_p)B = \eta_t\left( \nabla^2 - \frac{1}{s^2}  \right)B  
    + s\left(\left[ \nabla \times (A (r,\theta)\bf \hat{e}_\phi) \right]\cdot \nabla \Omega\right)
    + \frac{1}{s}\frac{\partial (sB)}{\partial r}\frac{\partial \eta_t}{\partial r},~~~
\end{equation}

\noindent where, $B (r, \theta)$ (i.e. $B_\phi$) and $A (r, \theta)$ are the toroidal and the poloidal (in the form of vector potential) components of the magnetic field respectively. Here $\Omega$ is the differential rotation, ${\bf v}_p$ is the meridional flow and $s = r\sin(\theta)$. This model presumes different diffusivity profiles for the toroidal and poloidal components of magnetic field: $\eta_t$ and $\eta_p$, respectively. In equation (10) `$\alpha B$' is the source term for generating B$_{p}$ and $\alpha$ incorporates contributions from both the B-L mechanism and mean field $\alpha$-effect. The details of every profile and parameter used in this model are elaborately described in an already published work\cite{2014A&A...563A..18P}. We note that no intrinsic amplitude fluctuation (over time) of the `$\alpha B$' is present in the model; the variation in poloidal field source term is being introduced only through the inclusion of the SFT results at every cycle minimum.

\begin{figure}[!ht]
\centering
\includegraphics[height=7cm, width=16cm]{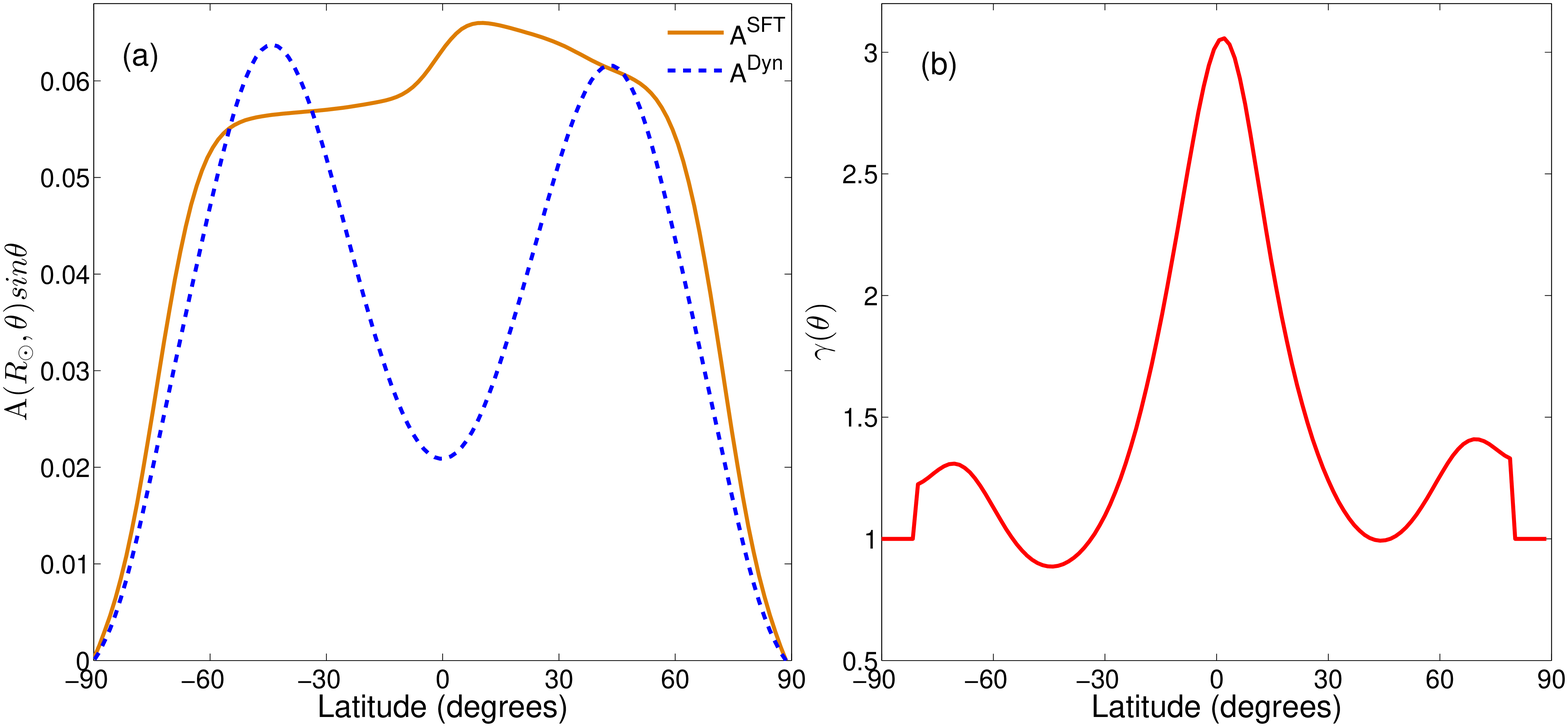}
\caption*{{\bf Fig. 5. Vector potential on the solar surface obtained from solar surface flux transport and dynamo simulations at the beginning of cycle 17.} (a) Depicts a comparison between A$^{SFT}(R_{\odot}, \theta) \sin \theta$ and A$^{Dyn}(R_{\odot}, \theta)\sin \theta$. (b) Depicts the associated $\gamma(\theta)$.}
\label{fig5}
\end{figure}

We incorporate results from SFT simulation into the dynamo model only during cycle minima, an approach similar to the earlier effort of predicting cycle 24 using observed surface magnetic field\cite{2007MNRAS.381.1527J}. The SFT model produces $B^{SFT}_r(R_{\odot},\theta, t_{min})$ which is related to $A^{SFT}(R_{\odot},\theta,t_{min})$ by

\begin{equation}
B^{SFT}_r(R_{\odot},\theta,t_{min})=\dfrac{1}{R_{\odot} \sin \theta} \dfrac{\partial}{\partial \theta} [\sin \theta A^{SFT}(R_{\odot},\theta,t_{min})]
\end{equation}

\noindent where we calculate B$^{SFT}_r$($R_{\odot}$, $\theta$, t$_{min}$) by averaging surface magnetic field over the $\phi$-direction during solar minima (t = t$_{min}$). We obtain A$^{SFT}$ on the surface for two hemispheres by using following relations,

\begin{equation}
  A^{SFT}(R_{\odot},\theta,t_{min}) \sin \theta =
  \begin{cases}
    \int_{0}^{\theta} B_r(R_{\odot},\theta^{\prime},t_{min}) \sin \theta^{\prime} d\theta^{\prime} \hspace{0.5cm} 0 < \theta < \pi / 2\\
   \int_{\pi}^{\theta} B_r(R_{\odot},\theta^{\prime},t_{min}) \sin \theta^{\prime} d\theta^{\prime} \hspace{0.5cm} \pi/2 < \theta < \pi.
  \end{cases}
\end{equation}

\noindent Now the dynamo generated $A^{Dyn}$ at solar minima is calibrated with the SFT generated A$^{SFT}$. We achieve these via two steps. In the first step we calibrate the amplitude of these two quantities at the solar surface by a factor ($c$), which once determined at the minimum of cycle 16, remains constant throughout our simulation. The A$^{Dyn}$($R_{\odot}$, $\theta$, t$_{min}$) $\sin \theta$ has different latitudinal distribution on the solar surface compared to A$^{SFT}$($R_{\odot}$, $\theta$, t$_{min}$) $\sin \theta$. In the second step of calibration, $A^{Dyn}$($R_{\odot}$, $\theta$, t$_{min}$) $\sin \theta$ obtained from the dynamo simulation at every minimum is multiplied by a function $\gamma(\theta)$ to make the product equal to A$^{SFT}$($R_{\odot}$, $\theta$, t$_{min}$) $\sin \theta$ on the solar surface. In Fig. 5a, the solid orange and the dashed blue lines depict the scaled A$^{SFT}$($R_{\odot}$, $\theta$) $\sin \theta$ and A$^{Dyn}$($R_{\odot}$, $\theta$) $\sin \theta$, respectively, at the minimum of cycle 16. Owing to the $\sin \theta$ term in the denominator, numerical issues arise while calculating the function $\gamma(\theta)$ near the polar regions. To avoid this, we set $\gamma(\theta)$ to be equal to one near the poles (see Fig. 5b). We assume that the poloidal field forcing (``correction'') on the dynamo due to the B-L mechanism is restricted to a region extending from 0.8 R$_{\odot}$ to R$_{\odot}$. During every solar minimum, we stop the dynamo simulation and multiply A$^{Dyn}$($r$, $\theta$, t$_{min}$) $\sin \theta$ with the appropriate $c \gamma(\theta)$  for every grid point above 0.8 R$_{\odot}$ and resume the simulation. At every subsequent minimum until cycle 24 minimum, this data assimilation from the SFT to the dynamo model is repeated; this generates our data-driven prediction for cycle 25. In spirit, this assimilation is akin to enforcing the data-driven SFT simulated surface map in the solar convection zone (the dynamo domain). Any transient discontinuities resulting from this ``driving'' is observed to disappear within a month, and the input is fully assimilated in the dynamo simulation.

The dynamo simulation provides a proxy for the toroidal magnetic field at the base of the SCZ which upon satisfying the magnetic buoyancy\cite{2014A&A...563A..18P,2002Sci...296.1671N,2004A&A...427.1019C} will appear as sunspots on the solar surface. We utilize the total erupted field,  B$^{Dyn}(t)$, as a proxy for total erupted sunspot flux (this is possible because each eruption have the same extent in radial and latitudinal grids). Therefore, we can compare B$^{Dyn}(t)$ with the unsigned flux associated with the observed sunspots (as depicted in Fig. 4). The dynamo simulated B$^{Dyn}(t)$ is calibrated with the observed unsigned magnetic flux through the utilization of a constant factor which remains the same throughout the simulation. This constant factor is determined through a multi-cycle (cycles 17--24) calibration of the annually averaged peak cycle strength (at each maximum) derived from the dynamo simulation, say $B_{max}(n)$ and the corresponding peak of the observed annually averaged, unsigned flux [$\Phi_{max}^{obs}(n)$]. We scale all $B_{max}(n)$ values corresponding to each maximum with the same constant factor, and vary the latter until the $B_{max}(n)$ versus $\Phi_{max}^{obs}(n)$ is characterized by a line with unit slope and zero intercept. We select this particular constant as the scaling factor which operates upon the whole B$^{Dyn}(t)$ time-series. The result is depicted in Fig. 4. A similar multi-cycle calibration technique is implemented to generate the amplitude prediction and range of the ensemble forecast in terms of the yearly mean sunspot number for cycle 25 (as reported in Table 1). This is achieved by calibrating the observed annually averaged peak sunspot numbers for cycles 17 to 24 maxima with the simulated peaks of the corresponding cycles.

At no point in our century-scale simulations is any individual scaling done to the amplitude of any single cycle, or any model driving parameters fine-tuned. This maintains the sanctity of these long-term data-driven simulations. 

\subsection*{Data availability}
The authors acknowledge utilization of data from the Royal Greenwich Observatory/USAF-NOAA active region database compiled by David H. Hathaway \url{(https://solarscience.msfc.nasa.gov/greenwch.shtml)}. MWO calibrated polar faculae data were downloaded from the solar dynamo database maintained by Andr\'{e}s Mu{\~n}oz-Jaramillo \url{(https://dataverse.harvard.edu/dataverse/solardynamo)}. The annual averaged sunspot number data is acquired from the World Data Center SILSO, Royal Observatory of Belgium, Brussels (\url{http://www.sidc.be/silso/datafiles}). The century-scale solar cycle simulation data and solar cycle 25 prediction data would be made available based on email requests to the corresponding author after a period of one year following publication.

\subsection*{Code availability} This work utilizes two disparate numerical codes for simulating magnetic field evolution on the solar surface and within the Sun's convection zone, respectively. Informed requests from established scientists for numerical simulations pertaining to this study may be entertained by the Center of Excellence in Space Sciences India. Such requests may be made through email to the corresponding author.

\vspace*{1cm}

\section*{Acknowledgements} 
This research was supported in parts by a CEFIPRA-IFCPAR grant 5004-1 and a NASA Heliophysics Grand Challenge grant NNX14AO83G. The Center of Excellence in Space Sciences India (\url{www.cessi.in}) -- where computer simulations related to this study were performed -- is funded by the Ministry of Human Resource Development, Government of India.

\section*{Author contributions}
The numerical modelling of the solar surface magnetic field evolution and solar internal dynamo simulations were performed by PB under the supervision of DN. Both authors contributed to discussions, preparation of results, and the writing of the manuscript.

\section*{Additional information}

\textbf{Competing interests:} The authors declare no competing interests.


\begin{thebibliography}{10}
\expandafter\ifx\csname url\endcsname\relax
  \def\url#1{\texttt{#1}}\fi
\expandafter\ifx\csname urlprefix\endcsname\relax\def\urlprefix{URL }\fi
\expandafter\ifx\csname doiprefix\endcsname\relax\def\doiprefix{DOI }\fi
\providecommand{\bibinfo}[2]{#2}
\providecommand{\eprint}[2][]{\url{#2}}

\bibitem{2015AdSpR..55.2745S}
\bibinfo{author}{{Schrijver}, C.~J.} \emph{et~al.}
\newblock \bibinfo{journal}{\bibinfo{title}{{Understanding space weather to
  shield society: A global road map for 2015-2025 commissioned by COSPAR and
  ILWS}}}.
\newblock {\emph{\JournalTitle{Advances in Space Research}}}
  \textbf{\bibinfo{volume}{55}}, \bibinfo{pages}{2745--2807}
  (\bibinfo{year}{2015}).
\newblock \doiprefix 10.1016/j.asr.2015.03.023.
\newblock \eprint{1503.06135}.

\bibitem{2005SSRv..120..243V}
\bibinfo{author}{{Versteegh}, G.~J.~M.}
\newblock \bibinfo{journal}{\bibinfo{title}{{Solar Forcing of Climate. 2:
  Evidence from the Past}}}.
\newblock {\emph{\JournalTitle{Space Sci. Rev.}}}
  \textbf{\bibinfo{volume}{120}}, \bibinfo{pages}{243--286}
  (\bibinfo{year}{2005}).
\newblock \doiprefix 10.1007/s11214-005-7047-4.

\bibitem{IPCC2014}
\bibinfo{author}{{IPCC 2014:}}.
\newblock \emph{\bibinfo{title}{{Climate Change 2014: Synthesis Report.
  Contribution of Working Groups I, II and III to the Fifth Assessment Report
  of the Intergovernmental Panel on Climate Change [Core Writing Team, R.K.
  Pachauri and L.A. Meyer (eds.)] }}}.
\newblock IPCC (\bibinfo{publisher}{Cambridge University Press},
  \bibinfo{address}{Geneva, Switzerland}, \bibinfo{year}{2014}).

\bibitem{1908ApJ....28..315H}
\bibinfo{author}{{Hale}, G.~E.}
\newblock \bibinfo{journal}{\bibinfo{title}{{On the probable existence of a
  magnetic field in Sun-spots}}}.
\newblock {\emph{\JournalTitle{Astrophys. J.}}} \textbf{\bibinfo{volume}{28}},
  \bibinfo{pages}{315} (\bibinfo{year}{1908}).
\newblock \doiprefix 10.1086/141602.

\bibitem{1919ApJ....49..153H}
\bibinfo{author}{{Hale}, G.~E.}, \bibinfo{author}{{Ellerman}, F.},
  \bibinfo{author}{{Nicholson}, S.~B.} \& \bibinfo{author}{{Joy}, A.~H.}
\newblock \bibinfo{journal}{\bibinfo{title}{{The magnetic polarity of
  Sun-spots}}}.
\newblock {\emph{\JournalTitle{Astrophys. J.}}} \textbf{\bibinfo{volume}{49}},
  \bibinfo{pages}{153} (\bibinfo{year}{1919}).
\newblock \doiprefix 10.1086/142452.

\bibitem{2010LRSP....7....3C}
\bibinfo{author}{{Charbonneau}, P.}
\newblock \bibinfo{journal}{\bibinfo{title}{{Dynamo models of the solar
  cycle}}}.
\newblock {\emph{\JournalTitle{Living Rev. Solar Phys.}}}
  \textbf{\bibinfo{volume}{7}}, \bibinfo{pages}{3} (\bibinfo{year}{2010}).
\newblock \doiprefix 10.12942/lrsp-2010-3.

\bibitem{1955ApJ...122..293P}
\bibinfo{author}{{Parker}, E.~N.}
\newblock \bibinfo{journal}{\bibinfo{title}{{Hydromagnetic dynamo models.}}}
\newblock {\emph{\JournalTitle{Astrophys. J.}}} \textbf{\bibinfo{volume}{122}},
  \bibinfo{pages}{293} (\bibinfo{year}{1955}).
\newblock \doiprefix 10.1086/146087.

\bibitem{1961ApJ...133..572B}
\bibinfo{author}{{Babcock}, H.~W.}
\newblock \bibinfo{journal}{\bibinfo{title}{{The topology of the Sun's magnetic
  field and the 22-year cycle.}}}
\newblock {\emph{\JournalTitle{Astrophys. J.}}} \textbf{\bibinfo{volume}{133}},
  \bibinfo{pages}{572} (\bibinfo{year}{1961}).
\newblock \doiprefix 10.1086/147060.

\bibitem{1969ApJ...156....1L}
\bibinfo{author}{{Leighton}, R.~B.}
\newblock \bibinfo{journal}{\bibinfo{title}{{A magneto-kinematic model of the
  solar cycle}}}.
\newblock {\emph{\JournalTitle{Astrophys. J.}}} \textbf{\bibinfo{volume}{156}},
  \bibinfo{pages}{1} (\bibinfo{year}{1969}).
\newblock \doiprefix 10.1086/149943.

\bibitem{2010A&A...518A...7D}
\bibinfo{author}{{Dasi-Espuig}, M.}, \bibinfo{author}{{Solanki}, S.~K.},
  \bibinfo{author}{{Krivova}, N.~A.}, \bibinfo{author}{{Cameron}, R.} \&
  \bibinfo{author}{{Pe{\~n}uela}, T.}
\newblock \bibinfo{journal}{\bibinfo{title}{{Sunspot group tilt angles and the
  strength of the solar cycle}}}.
\newblock {\emph{\JournalTitle{Astron. Astrophys.}}}
  \textbf{\bibinfo{volume}{518}}, \bibinfo{pages}{A7} (\bibinfo{year}{2010}).
\newblock \doiprefix 10.1051/0004-6361/201014301.
\newblock \eprint{1005.1774}.

\bibitem{2015Sci...347.1333C}
\bibinfo{author}{{Cameron}, R.} \& \bibinfo{author}{{Sch{\"u}ssler}, M.}
\newblock \bibinfo{journal}{\bibinfo{title}{{The crucial role of surface
  magnetic fields for the solar dynamo}}}.
\newblock {\emph{\JournalTitle{Science}}} \textbf{\bibinfo{volume}{347}},
  \bibinfo{pages}{1333--1335} (\bibinfo{year}{2015}).
\newblock \doiprefix 10.1126/science.1261470.
\newblock \eprint{1503.08469}.

\bibitem{2007ApJ...661.1289B}
\bibinfo{author}{{Bushby}, P.~J.} \& \bibinfo{author}{{Tobias}, S.~M.}
\newblock \bibinfo{journal}{\bibinfo{title}{{On predicting the solar cycle
  Using mean-field models}}}.
\newblock {\emph{\JournalTitle{Astrophys. J.}}} \textbf{\bibinfo{volume}{661}},
  \bibinfo{pages}{1289--1296} (\bibinfo{year}{2007}).
\newblock \doiprefix 10.1086/516628.
\newblock \eprint{0704.2345}.

\bibitem{2008SoPh..252..209P}
\bibinfo{author}{{Pesnell}, W.~D.}
\newblock \bibinfo{journal}{\bibinfo{title}{{Predictions of solar cycle 24}}}.
\newblock {\emph{\JournalTitle{Sol. Phys.}}} \textbf{\bibinfo{volume}{252}},
  \bibinfo{pages}{209--220} (\bibinfo{year}{2008}).
\newblock \doiprefix 10.1007/s11207-008-9252-2.

\bibitem{2006GeoRL..33.5102D}
\bibinfo{author}{{Dikpati}, M.}, \bibinfo{author}{{de Toma}, G.} \&
  \bibinfo{author}{{Gilman}, P.~A.}
\newblock \bibinfo{journal}{\bibinfo{title}{{Predicting the strength of solar
  cycle 24 using a flux-transport dynamo-based tool}}}.
\newblock {\emph{\JournalTitle{GeoRL}}} \textbf{\bibinfo{volume}{33}},
  \bibinfo{pages}{L05102} (\bibinfo{year}{2006}).
\newblock \doiprefix 10.1029/2005GL025221.

\bibitem{2007PhRvL..98m1103C}
\bibinfo{author}{{Choudhuri}, A.~R.}, \bibinfo{author}{{Chatterjee}, P.} \&
  \bibinfo{author}{{Jiang}, J.}
\newblock \bibinfo{journal}{\bibinfo{title}{{Predicting solar cycle 24 with a
  solar dynamo model}}}.
\newblock {\emph{\JournalTitle{Physical Review Letters}}}
  \textbf{\bibinfo{volume}{98}}, \bibinfo{pages}{131103}
  (\bibinfo{year}{2007}).
\newblock \doiprefix 10.1103/PhysRevLett.98.131103.
\newblock \eprint{astro-ph/0701527}.
\bibitem{2007MNRAS.381.1527J}
\bibinfo{author}{{Jiang}, J.}, \bibinfo{author}{{Chatterjee}, P.} \&
  \bibinfo{author}{{Choudhuri}, A.~R.}
\newblock \bibinfo{journal}{\bibinfo{title}{{Solar activity forecast with a
  dynamo model}}}.
\newblock {\emph{\JournalTitle{MNRAS}}} \textbf{\bibinfo{volume}{381}},
  \bibinfo{pages}{1527--1542} (\bibinfo{year}{2007}).
\newblock \doiprefix 10.1111/j.1365-2966.2007.12267.x.
\newblock \eprint{0707.2258}.


\bibitem{2008ApJ...673..544Y}
\bibinfo{author}{{Yeates}, A.~R.}, \bibinfo{author}{{Nandy}, D.} \&
  \bibinfo{author}{{Mackay}, D.~H.}
\newblock \bibinfo{journal}{\bibinfo{title}{{Exploring the physical basis of
  solar cycle predictions: flux transport dynamics and persistence of memory in
  advection- versus diffusion-dominated solar convection zones}}}.
\newblock {\emph{\JournalTitle{Astrophys. J.}}} \textbf{\bibinfo{volume}{673}},
  \bibinfo{pages}{544--556} (\bibinfo{year}{2008}).
\newblock \doiprefix 10.1086/524352.
\newblock \eprint{0709.1046}.

\bibitem{2010LRSP....7....6P}
\bibinfo{author}{{Petrovay}, K.}
\newblock \bibinfo{journal}{\bibinfo{title}{{Solar cycle prediction}}}.
\newblock {\emph{\JournalTitle{Living Rev. Solar Phys.}}}
  \textbf{\bibinfo{volume}{7}}, \bibinfo{pages}{6} (\bibinfo{year}{2010}).
\newblock \doiprefix 10.12942/lrsp-2010-6.
\newblock \eprint{1012.5513}.

\bibitem{2012ApJ...761L..13K}
\bibinfo{author}{{Karak}, B.~B.} \& \bibinfo{author}{{Nandy}, D.}
\newblock \bibinfo{journal}{\bibinfo{title}{{Turbulent pumping of magnetic flux
  reduces solar cycle memory and thus impacts predictability of the Sun's
  activity}}}.
\newblock {\emph{\JournalTitle{Astrophys. J. Lett.}}}
  \textbf{\bibinfo{volume}{761}}, \bibinfo{pages}{L13} (\bibinfo{year}{2012}).
\newblock \doiprefix 10.1088/2041-8205/761/1/L13.
\newblock \eprint{1206.2106}.

\bibitem{1999JGR...10422375H}
\bibinfo{author}{{Hathaway}, D.~H.}, \bibinfo{author}{{Wilson}, R.~M.} \&
  \bibinfo{author}{{Reichmann}, E.~J.}
\newblock \bibinfo{journal}{\bibinfo{title}{{A synthesis of solar cycle
  prediction techniques}}}.
\newblock {\emph{\JournalTitle{J. Geophys. Res.}}}
  \textbf{\bibinfo{volume}{104}}, \bibinfo{pages}{22} (\bibinfo{year}{1999}).
\newblock \doiprefix 10.1029/1999JA900313.

\bibitem{2005GeoRL..32.1104S}
\bibinfo{author}{{Svalgaard}, L.}, \bibinfo{author}{{Cliver}, E.~W.} \&
  \bibinfo{author}{{Kamide}, Y.}
\newblock \bibinfo{journal}{\bibinfo{title}{{Sunspot cycle 24: smallest cycle
  in 100 years?}}}
\newblock {\emph{\JournalTitle{GeoRL}}} \textbf{\bibinfo{volume}{32}},
  \bibinfo{pages}{L01104} (\bibinfo{year}{2005}).
\newblock \doiprefix 10.1029/2004GL021664.

\bibitem{2009ApJ...694L..11W}
\bibinfo{author}{{Wang}, Y.-M.} \& \bibinfo{author}{{Sheeley}, N.~R.}
\newblock \bibinfo{journal}{\bibinfo{title}{{Understanding the geomagnetic
  precursor of the solar cycle}}}.
\newblock {\emph{\JournalTitle{Astrophys. J. Lett.}}}
  \textbf{\bibinfo{volume}{694}}, \bibinfo{pages}{L11--L15}
  (\bibinfo{year}{2009}).
\newblock \doiprefix 10.1088/0004-637X/694/1/L11.

\bibitem{2013ApJ...767L..25M}
\bibinfo{author}{{Mu{\~n}oz-Jaramillo}, A.}, \bibinfo{author}{{Dasi-Espuig},
  M.}, \bibinfo{author}{{Balmaceda}, L.~A.} \& \bibinfo{author}{{DeLuca},
  E.~E.}
\newblock \bibinfo{journal}{\bibinfo{title}{{Solar cycle propagation, memory,
  and prediction: insights from a century of magnetic proxies}}}.
\newblock {\emph{\JournalTitle{Astrophys. J. Lett.}}}
  \textbf{\bibinfo{volume}{767}}, \bibinfo{pages}{L25} (\bibinfo{year}{2013}).
\newblock \doiprefix 10.1088/2041-8205/767/2/L25.
\newblock \eprint{1304.3151}.

\bibitem{1989Sci...245..712W}
\bibinfo{author}{{Wang}, Y.-M.}, \bibinfo{author}{{Nash}, A.~G.} \&
  \bibinfo{author}{{Sheeley}, N.~R., Jr.}
\newblock \bibinfo{journal}{\bibinfo{title}{{Magnetic flux transport on the
  sun}}}.
\newblock {\emph{\JournalTitle{Science}}} \textbf{\bibinfo{volume}{245}},
  \bibinfo{pages}{712--718} (\bibinfo{year}{1989}).
\newblock \doiprefix 10.1126/science.245.4919.712.

\bibitem{1998ApJ...501..866V}
\bibinfo{author}{{van Ballegooijen}, A.~A.}, \bibinfo{author}{{Cartledge},
  N.~P.} \& \bibinfo{author}{{Priest}, E.~R.}
\newblock \bibinfo{journal}{\bibinfo{title}{{Magnetic flux transport and the
  formation of filament channels on the sun}}}.
\newblock {\emph{\JournalTitle{Astrophys. J.}}} \textbf{\bibinfo{volume}{501}},
  \bibinfo{pages}{866--881} (\bibinfo{year}{1998}).
\newblock \doiprefix 10.1086/305823.

\bibitem{2001ApJ...547..475S}
\bibinfo{author}{{Schrijver}, C.~J.}
\newblock \bibinfo{journal}{\bibinfo{title}{{Simulations of the Photospheric
  Magnetic Activity and Outer Atmospheric Radiative Losses of Cool Stars Based
  on Characteristics of the Solar Magnetic Field}}}.
\newblock {\emph{\JournalTitle{Astrophys. J.}}} \textbf{\bibinfo{volume}{547}},
  \bibinfo{pages}{475--490} (\bibinfo{year}{2001}).
\newblock \doiprefix 10.1086/318333.

\bibitem{2005LRSP....2....5S}
\bibinfo{author}{{Sheeley}, N.~R., Jr.}
\newblock \bibinfo{journal}{\bibinfo{title}{{Surface evolution of the Sun's
  magnetic field: A historical review of the flux-transport mechanism}}}.
\newblock {\emph{\JournalTitle{Living Rev. Solar Phys.}}}
  \textbf{\bibinfo{volume}{2}}, \bibinfo{pages}{5} (\bibinfo{year}{2005}).
\newblock \doiprefix 10.12942/lrsp-2005-5.

\bibitem{2010ApJ...719..264C}
\bibinfo{author}{{Cameron}, R.~H.}, \bibinfo{author}{{Jiang}, J.},
  \bibinfo{author}{{Schmitt}, D.} \& \bibinfo{author}{{Sch{\"u}ssler}, M.}
\newblock \bibinfo{journal}{\bibinfo{title}{{Surface flux Transport Modeling
  for Solar Cycles 15-21: effects of cycle-dependent tilt angles of sunspot
  groups}}}.
\newblock {\emph{\JournalTitle{Astrophys. J.}}} \textbf{\bibinfo{volume}{719}},
  \bibinfo{pages}{264--270} (\bibinfo{year}{2010}).
\newblock \doiprefix 10.1088/0004-637X/719/1/264.
\newblock \eprint{1006.3061}.

\bibitem{2012LRSP....9....6M}
\bibinfo{author}{{Mackay}, D.~H.} \& \bibinfo{author}{{Yeates}, A.~R.}
\newblock \bibinfo{journal}{\bibinfo{title}{{The Sun's global photospheric and
  coronal magnetic fields: observations and models}}}.
\newblock {\emph{\JournalTitle{Living Rev. Solar Phys.}}}
  \textbf{\bibinfo{volume}{9}}, \bibinfo{pages}{6} (\bibinfo{year}{2012}).
\newblock \doiprefix 10.12942/lrsp-2012-6.
\newblock \eprint{1211.6545}.

\bibitem{2014SSRv..186..491J}
\bibinfo{author}{{Jiang}, J.} \emph{et~al.}
\newblock \bibinfo{journal}{\bibinfo{title}{{Magnetic Flux Transport at the
  Solar Surface}}}.
\newblock {\emph{\JournalTitle{Space Sci. Rev.}}}
  \textbf{\bibinfo{volume}{186}}, \bibinfo{pages}{491--523}
  (\bibinfo{year}{2014}).
\newblock \doiprefix 10.1007/s11214-014-0083-1.
\newblock \eprint{1408.3186}.

\bibitem{2016JGRA..12110744H}
\bibinfo{author}{{Hathaway}, D.~H.} \& \bibinfo{author}{{Upton}, L.~A.}
\newblock \bibinfo{journal}{\bibinfo{title}{{Predicting the amplitude and
  hemispheric asymmetry of solar cycle 25 with surface flux transport}}}.
\newblock {\emph{\JournalTitle{Journal of Geophysical Research (Space
  Physics)}}} \textbf{\bibinfo{volume}{121}}, \bibinfo{pages}{10}
  (\bibinfo{year}{2016}).
\newblock \doiprefix 10.1002/2016JA023190.
\newblock \eprint{1611.05106}.

\bibitem{2016ApJ...823L..22C}
\bibinfo{author}{{Cameron}, R.~H.}, \bibinfo{author}{{Jiang}, J.} \&
  \bibinfo{author}{{Sch{\"u}ssler}, M.}
\newblock \bibinfo{journal}{\bibinfo{title}{{Solar cycle 25: another moderate
  cycle?}}}
\newblock {\emph{\JournalTitle{Astrophys. J. Lett.}}}
  \textbf{\bibinfo{volume}{823}}, \bibinfo{pages}{L22} (\bibinfo{year}{2016}).
\newblock \doiprefix 10.3847/2041-8205/823/2/L22.
\newblock \eprint{1604.05405}.

\bibitem{2012ApJ...753..146M}
\bibinfo{author}{{Mu{\~n}oz-Jaramillo}, A.}, \bibinfo{author}{{Sheeley},
  N.~R.}, \bibinfo{author}{{Zhang}, J.} \& \bibinfo{author}{{DeLuca}, E.~E.}
\newblock \bibinfo{journal}{\bibinfo{title}{{Calibrating 10 years of polar
  faculae measurements: implications for the evolution of the heliospheric
  magnetic field}}}.
\newblock {\emph{\JournalTitle{Astrophys. J.}}} \textbf{\bibinfo{volume}{753}},
  \bibinfo{pages}{146} (\bibinfo{year}{2012}).
\newblock \doiprefix 10.1088/0004-637X/753/2/146.
\newblock \eprint{1303.0345}.

\bibitem{1999SoPh..189...69S}
\bibinfo{author}{{Sivaraman}, K.~R.}, \bibinfo{author}{{Gupta}, S.~S.} \&
  \bibinfo{author}{{Howard}, R.~F.}
\newblock \bibinfo{journal}{\bibinfo{title}{{Measurement of Kodaikanal
  white-light images - IV. axial tilt angles of sunspot groups}}}.
\newblock {\emph{\JournalTitle{Sol. Phys.}}} \textbf{\bibinfo{volume}{189}},
  \bibinfo{pages}{69--83} (\bibinfo{year}{1999}).
\newblock \doiprefix 10.1023/A:1005277515551.

\bibitem{2013SoPh..287..215M}
\bibinfo{author}{{McClintock}, B.~H.} \& \bibinfo{author}{{Norton}, A.~A.}
\newblock \bibinfo{journal}{\bibinfo{title}{{Recovering Joy's law as a function
  of solar cycle, hemisphere, and longitude}}}.
\newblock {\emph{\JournalTitle{Sol. Phys.}}} \textbf{\bibinfo{volume}{287}},
  \bibinfo{pages}{215--227} (\bibinfo{year}{2013}).
\newblock \doiprefix 10.1007/s11207-013-0338-0.
\newblock \eprint{1305.3205}.

\bibitem{2014ApJ...791....5J}
\bibinfo{author}{{Jiang}, J.}, \bibinfo{author}{{Cameron}, R.~H.} \&
  \bibinfo{author}{{Sch{\"u}ssler}, M.}
\newblock \bibinfo{journal}{\bibinfo{title}{{Effects of the scatter in sunspot
  group tilt angles on the large-scale magnetic field at the solar surface}}}.
\newblock {\emph{\JournalTitle{Astrophys. J.}}} \textbf{\bibinfo{volume}{791}},
  \bibinfo{pages}{5} (\bibinfo{year}{2014}).
\newblock \doiprefix 10.1088/0004-637X/791/1/5.
\newblock \eprint{1406.5564}.

\bibitem{2017SoPh..292..167N}
\bibinfo{author}{{Nagy}, M.}, \bibinfo{author}{{Lemerle}, A.},
  \bibinfo{author}{{Labonville}, F.}, \bibinfo{author}{{Petrovay}, K.} \&
  \bibinfo{author}{{Charbonneau}, P.}
\newblock \bibinfo{journal}{\bibinfo{title}{{The Effect of ``Rogue'' Active
  Regions on the Solar Cycle}}}.
\newblock {\emph{\JournalTitle{Sol. Phys.}}} \textbf{\bibinfo{volume}{292}},
  \bibinfo{pages}{167} (\bibinfo{year}{2017}).
\newblock \doiprefix 10.1007/s11207-017-1194-0.
\newblock \eprint{1712.02185}.

\bibitem{2015SoPh..290.3189Y}
\bibinfo{author}{{Yeates}, A.~R.}, \bibinfo{author}{{Baker}, D.} \&
  \bibinfo{author}{{van Driel-Gesztelyi}, L.}
\newblock \bibinfo{journal}{\bibinfo{title}{{Source of a prominent poleward
  surge during solar cycle 24}}}.
\newblock {\emph{\JournalTitle{Sol. Phys.}}} \textbf{\bibinfo{volume}{290}},
  \bibinfo{pages}{3189--3201} (\bibinfo{year}{2015}).
\newblock \doiprefix 10.1007/s11207-015-0660-9.
\newblock \eprint{1502.04854}.

\bibitem{2014ApJ...792..142U}
\bibinfo{author}{{Upton}, L.} \& \bibinfo{author}{{Hathaway}, D.~H.}
\newblock \bibinfo{journal}{\bibinfo{title}{{Effects of Meridional Flow
  Variations on Solar Cycles 23 and 24}}}.
\newblock {\emph{\JournalTitle{Astrophys. J.}}} \textbf{\bibinfo{volume}{792}},
  \bibinfo{pages}{142} (\bibinfo{year}{2014}).
\newblock \doiprefix 10.1088/0004-637X/792/2/142.
\newblock \eprint{1408.0035}.

\bibitem{2014A&A...563A..18P}
\bibinfo{author}{{Passos}, D.}, \bibinfo{author}{{Nandy}, D.},
  \bibinfo{author}{{Hazra}, S.} \& \bibinfo{author}{{Lopes}, I.}
\newblock \bibinfo{journal}{\bibinfo{title}{{A solar dynamo model driven by
  mean-field alpha and Babcock-Leighton sources: fluctuations,
  grand-minima-maxima, and hemispheric asymmetry in sunspot cycles}}}.
\newblock {\emph{\JournalTitle{Astron. Astrophys.}}}
  \textbf{\bibinfo{volume}{563}}, \bibinfo{pages}{A18} (\bibinfo{year}{2014}).
\newblock \doiprefix 10.1051/0004-6361/201322635.
\newblock \eprint{1309.2186}.

\bibitem{2017ApJ...834..133L}
\bibinfo{author}{{Lemerle}, A.} \& \bibinfo{author}{{Charbonneau}, P.}
\newblock \bibinfo{journal}{\bibinfo{title}{{A coupled 2 ${\times}$ 2D
  Babcock-Leighton solar dynamo model. II. reference dynamo solutions}}}.
\newblock {\emph{\JournalTitle{Astrophys. J.}}} \textbf{\bibinfo{volume}{834}},
  \bibinfo{pages}{133} (\bibinfo{year}{2017}).
\newblock \doiprefix 10.3847/1538-4357/834/2/133.
\newblock \eprint{1606.07375}.

\bibitem{2002Sci...296.1671N}
\bibinfo{author}{{Nandy}, D.} \& \bibinfo{author}{{Choudhuri}, A.~R.}
\newblock \bibinfo{journal}{\bibinfo{title}{{Explaining the latitudinal
  distribution of sunspots with deep meridional flow}}}.
\newblock {\emph{\JournalTitle{Science}}} \textbf{\bibinfo{volume}{296}},
  \bibinfo{pages}{1671--1673} (\bibinfo{year}{2002}).
\newblock \doiprefix 10.1126/science.1070955.

\bibitem{2004A&A...427.1019C}
\bibinfo{author}{{Chatterjee}, P.}, \bibinfo{author}{{Nandy}, D.} \&
  \bibinfo{author}{{Choudhuri}, A.~R.}
\newblock \bibinfo{journal}{\bibinfo{title}{{Full-sphere simulations of a
  circulation-dominated solar dynamo: Exploring the parity issue}}}.
\newblock {\emph{\JournalTitle{Astron. Astrophys.}}}
  \textbf{\bibinfo{volume}{427}}, \bibinfo{pages}{1019--1030}
  (\bibinfo{year}{2004}).
\newblock \doiprefix 10.1051/0004-6361:20041199.
\newblock \eprint{astro-ph/0405027}.

\bibitem{2002SoPh..211..357H}
\bibinfo{author}{{Hathaway}, D.~H.}, \bibinfo{author}{{Wilson}, R.~M.} \&
  \bibinfo{author}{{Reichmann}, E.~J.}
\newblock \bibinfo{journal}{\bibinfo{title}{{Group Sunspot Numbers: Sunspot
  Cycle Characteristics}}}.
\newblock {\emph{\JournalTitle{Sol. Phys.}}} \textbf{\bibinfo{volume}{211}},
  \bibinfo{pages}{357--370} (\bibinfo{year}{2002}).
\newblock \doiprefix 10.1023/A:1022425402664.

\bibitem{2014ApJ...789....5H}
\bibinfo{author}{{Hazra}, S.}, \bibinfo{author}{{Passos}, D.} \&
  \bibinfo{author}{{Nandy}, D.}
\newblock \bibinfo{journal}{\bibinfo{title}{{A stochastically forced time delay
  solar dynamo model: self-consistent recovery from a Maunder-like grand
  minimum necessitates a mean-field alpha effect}}}.
\newblock {\emph{\JournalTitle{Astrophys. J.}}} \textbf{\bibinfo{volume}{789}},
  \bibinfo{pages}{5} (\bibinfo{year}{2014}).
\newblock \doiprefix 10.1088/0004-637X/789/1/5.
\newblock \eprint{1307.5751}.

\bibitem{2018ApJ...863..159J}
\bibinfo{author}{{Jiang}, J.}, \bibinfo{author}{{Wang}, J.~X.} \&
  \bibinfo{author}{{Jiao}, Q~R.}, \bibinfo{author}{{Cao}, J.~B.}
\newblock \bibinfo{journal}{\bibinfo{title}{{Predictability of the Solar Cycle Over One Cycle}}}.
\newblock {\emph{\JournalTitle{Astrophys. J.}}} \textbf{\bibinfo{volume}{863}},
  \bibinfo{pages}{159} (\bibinfo{year}{2018}).
\newblock \doiprefix 10.3847/1538-4357/aad197.
\newblock \eprint{1807.01543}.

\bibitem{1993SSRv...63....1S}
\bibinfo{author}{{Solanki}, S.~K.}
\newblock \bibinfo{journal}{\bibinfo{title}{{Small-scale solar magnetic fields
  - an overview}}}.
\newblock {\emph{\JournalTitle{Space Sci. Rev.}}}
  \textbf{\bibinfo{volume}{63}}, \bibinfo{pages}{1--188}
  (\bibinfo{year}{1993}).
\newblock \doiprefix 10.1007/BF00749277.

\bibitem{1983ApJ...270..288S}
\bibinfo{author}{{Snodgrass}, H.~B.}
\newblock \bibinfo{journal}{\bibinfo{title}{{Magnetic rotation of the solar
  photosphere}}}.
\newblock {\emph{\JournalTitle{Astrophys. J.}}} \textbf{\bibinfo{volume}{270}},
  \bibinfo{pages}{288--299} (\bibinfo{year}{1983}).
\newblock \doiprefix 10.1086/161121.

\bibitem{1998ApJ...505..390S}
\bibinfo{author}{{Schou}, J.} \emph{et~al.}
\newblock \bibinfo{journal}{\bibinfo{title}{{Helioseismic studies of
  differential rotation in the solar envelope by the solar oscillations
  investigation using the Michelson Doppler Imager}}}.
\newblock {\emph{\JournalTitle{Astrophys. J.}}} \textbf{\bibinfo{volume}{505}},
  \bibinfo{pages}{390--417} (\bibinfo{year}{1998}).
\newblock \doiprefix 10.1086/306146.

\bibitem{1994ApJ...430..399W}
\bibinfo{author}{{Wang}, Y.-M.} \& \bibinfo{author}{{Sheeley}, N.~R., Jr.}
\newblock \bibinfo{journal}{\bibinfo{title}{{The rotation of photospheric
  magnetic fields: A random walk transport model}}}.
\newblock {\emph{\JournalTitle{Astrophys. J.}}} \textbf{\bibinfo{volume}{430}},
  \bibinfo{pages}{399--412} (\bibinfo{year}{1994}).
\newblock \doiprefix 10.1086/174415.

\bibitem{2000ssma.book.....S}
\bibinfo{author}{{Schrijver}, C.~J.} \& \bibinfo{author}{{Zwaan}, C.}
\newblock \emph{\bibinfo{title}{{Solar and Stellar Magnetic Activity}}},
  vol.~\bibinfo{volume}{34} of \emph{\bibinfo{series}{Cambridge astrophysics
  series}} (\bibinfo{publisher}{Cambridge University Press},
  \bibinfo{address}{New York}, \bibinfo{year}{2000}).

\bibitem{2002ApJ...577L..53W}
\bibinfo{author}{{Wang}, Y.-M.}, \bibinfo{author}{{Lean}, J.} \&
  \bibinfo{author}{{Sheeley}, N.~R., Jr.}
\newblock \bibinfo{journal}{\bibinfo{title}{{Role of a variable meridional flow
  in the secular evolution of the Sun's polar fields and open flux}}}.
\newblock {\emph{\JournalTitle{Astrophys. J. Lett.}}}
  \textbf{\bibinfo{volume}{577}}, \bibinfo{pages}{L53--L57}
  (\bibinfo{year}{2002}).
\newblock \doiprefix 10.1086/344196.

\bibitem{2002ApJ...577.1006S}
\bibinfo{author}{{Schrijver}, C.~J.}, \bibinfo{author}{{De Rosa}, M.~L.} \&
  \bibinfo{author}{{Title}, A.~M.}
\newblock \bibinfo{journal}{\bibinfo{title}{{What is missing from Our
  understanding of long-term solar and heliospheric activity?}}}
\newblock {\emph{\JournalTitle{Astrophys. J.}}} \textbf{\bibinfo{volume}{577}},
  \bibinfo{pages}{1006--1012} (\bibinfo{year}{2002}).
\newblock \doiprefix 10.1086/342247.

\bibitem{2006A&A...446..307B}
\bibinfo{author}{{Baumann}, I.}, \bibinfo{author}{{Schmitt}, D.} \&
  \bibinfo{author}{{Sch{\"u}ssler}, M.}
\newblock \bibinfo{journal}{\bibinfo{title}{{A necessary extension of the
  surface flux transport model}}}.
\newblock {\emph{\JournalTitle{Astron. Astrophys.}}}
  \textbf{\bibinfo{volume}{446}}, \bibinfo{pages}{307--314}
  (\bibinfo{year}{2006}).
\newblock \doiprefix 10.1051/0004-6361:20053488.

\bibitem{2015ApJ...800...48M}
\bibinfo{author}{{Mu{\~n}oz-Jaramillo}, A.} \emph{et~al.}
\newblock \bibinfo{journal}{\bibinfo{title}{{Small-scale and global dynamos and
  the area and flux distributions of active regions, sunspot groups, and
  sunspots: a multi-database study}}}.
\newblock {\emph{\JournalTitle{Astrophys. J.}}} \textbf{\bibinfo{volume}{800}},
  \bibinfo{pages}{48} (\bibinfo{year}{2015}).
\newblock \doiprefix 10.1088/0004-637X/800/1/48.
\newblock \eprint{1410.6281}.

\bibitem{2014SoPh..289.1517F}
\bibinfo{author}{{Foukal}, P.}
\newblock \bibinfo{journal}{\bibinfo{title}{{An Explanation of the Differences
  Between the Sunspot Area Scales of the Royal Greenwich and Mt. Wilson
  Observatories, and the SOON Program}}}.
\newblock {\emph{\JournalTitle{Sol. Phys.}}} \textbf{\bibinfo{volume}{289}},
  \bibinfo{pages}{1517--1529} (\bibinfo{year}{2014}).
\newblock \doiprefix 10.1007/s11207-013-0425-2.

\bibitem{2011A&A...528A..82J}
\bibinfo{author}{{Jiang}, J.}, \bibinfo{author}{{Cameron}, R.~H.},
  \bibinfo{author}{{Schmitt}, D.} \& \bibinfo{author}{{Sch{\"u}ssler}, M.}
\newblock \bibinfo{journal}{\bibinfo{title}{{The solar magnetic field since
  1700. I. Characteristics of sunspot group emergence and reconstruction of the
  butterfly diagram}}}.
\newblock {\emph{\JournalTitle{Astron. Astrophys.}}}
  \textbf{\bibinfo{volume}{528}}, \bibinfo{pages}{A82} (\bibinfo{year}{2011}).
\newblock \doiprefix 10.1051/0004-6361/201016167.
\newblock \eprint{1102.1266}.

\bibitem{1994SoPh..151..177H}
\bibinfo{author}{{Hathaway}, D.~H.}, \bibinfo{author}{{Wilson}, R.~M.} \&
  \bibinfo{author}{{Reichmann}, E.~J.}
\newblock \bibinfo{journal}{\bibinfo{title}{{The shape of the sunspot cycle}}}.
\newblock {\emph{\JournalTitle{Sol. Phys.}}} \textbf{\bibinfo{volume}{151}},
  \bibinfo{pages}{177--190} (\bibinfo{year}{1994}).
\newblock \doiprefix 10.1007/BF00654090.

\bibitem{1999ApJ...518..508D}
\bibinfo{author}{{Dikpati}, M.} \& \bibinfo{author}{{Charbonneau}, P.}
\newblock \bibinfo{journal}{\bibinfo{title}{{A Babcock-Leighton Flux Transport
  Dynamo with Solar-like Differential Rotation}}}.
\newblock {\emph{\JournalTitle{Astrophys. J.}}} \textbf{\bibinfo{volume}{518}},
  \bibinfo{pages}{508--520} (\bibinfo{year}{1999}).
\newblock \doiprefix 10.1086/307269.

\end{thebibliography}
\end{document}